\documentclass[%
preprint,
amsmath,amssymb,
prc,
]{revtex4-1}

\usepackage[dvipdfmx]{graphicx}
\usepackage{dcolumn}
\usepackage{bm}
\usepackage{braket}
\usepackage{color}
\usepackage{here}

\begin{document}
\title{
Continuum random-phase approximation for $(n,\gamma)$ reactions on neutron-rich nuclei: 
collective effects and resonances
}
\author{Teruyuki Saito}
 \affiliation{Graduate School of Science and Technology, Niigata University, Niigata 950-2181, Japan}
\author{Masayuki Matsuo}
\affiliation{Department of Physics, Faculty of Science. Niigata University, Niigata 950-2181, Japan}

\begin{abstract}
We formulate a microscopic theory
to calculate cross section of the radiative neutron capture
reaction on neutron-rich nuclei using the continuum random-phase approximation
(cRPA) to the time-dependent density functional theory (TDDFT). 
With an intension of applying to the r-process, for which the statistical reaction model may not be appropriate,
we describe the transition between a initial state of incident neutron and  
a final state of the gamma decay by means of a single many-body framework of the cRPA-TDDFT. 
With the cRPA approach, it is possible to describe various excitation modes present in the $(n,\gamma)$ reaction,
 including soft dipole excitation, the giant resonances as well as  non-collective excitations and
the single-particle resonances. Furthermore, it enables us to describe the $(n,\gamma)$ reaction where the 
the final states of the gamma transition are low-lying surface vibrational states.
We demonstrate the theory by performing
numerical calculation for the reaction $^{139}{\rm Sn} (n,\gamma) ^{140}{\rm Sn}$.
We discuss various new features which are beyond the
single-particle model; presence of narrow and wide resonances originating 
from non-collective and collective excitations and roles of low-lying quadrupole and octupole 
vibrational states.
\end{abstract}

\maketitle

\section{Introduction}
The r-process  is believed to be the origin of about half  of the elements heavier than iron \cite{B2FH,Arnould2007,Cowan2021}.
From the viewpoint of nuclear physics, the nucleosynthesis is
a complex network of the  $(n,\gamma)$ and $(\gamma,n)$ reactions, the beta decay and the fission
taking place on  a large number of short-lived neutron-rich nuclei very far from the stability
line. 
The cross sections and the probabilities of the relevant reactions and decays need to be
provided theoretically for quantitative understanding since the direct experimental measurements are quite difficult or impossible 
in most cases.
Recent observation of kilonova associated with a binary neutron star merger \cite{Abbott2017, Abbott2017_2} provides a first direct evidence 
and quantitative information on the r-process nucleosynthesis \cite{Pian2017,Kasen2017,Villar2017,Cowan2021}. 
It is therefore quite important for nuclear theory
to provide reliable theoretical models of the relevant nuclear reactions which takes into account the
recent progress of study  of neutron-rich nuclei.

In the present study, we focus on the radiative neutron capture reactions.
It is usually described in terms of two different mechanisms, the compound nuclear (CN) process and direct capture (DC) process \cite{Arnould2007, Cowan2021}.
The CN process is assumed to proceed via compound states with high level density,
and it is usually evaluated by means of the Hauser and Feshbach statistical model \cite{Hauser1952}. It
 is relevant for  nuclei with relatively large neutron separation energy and 
 and often applied to the s-process which occurs on stable nuclei or nuclei close to the
stability line.  The r-process path, however,  lies on short-lived neutron-rich nuclei in which the
neutron separation energy or the excitation energy of nuclei is as small as $\sim 2$ MeV, In this case
the statistical description of the CN process may not be appropriate \cite{Mathews1983,Arnould2007, Cowan2021}. 
Often considered in this case is the direct capture process \cite{Lane1960, Mathews1983, Raman1985, Mengoni1995, Rauscher1998, Bonneau2007, Chiba2008, Rauscher2010, Xu2012, Xu2014, Zhang2015, Sieja2021}, where a neutron scattering state decays 
directly to a bound state without forming compound states. It is essentially a single-particle model of
gamma decay or an independent-particle shell model for the nuclear structure.

There exists various nuclear structure phenomena beyond the single-particle description, such as
the pairing correlation, the low-lying  collective states, and giant resonances. 
In addition, 
neutron-rich nuclei far from the stability line exhibits 
characteristic features such as the neutron halo,the neutron skin and related exotic modes such as the pygmy dipole resonance or the soft dipole mode \cite{Tanihata1985, Hansen1987, Suzuki1990, Bertsch1991, Paar2007, Tanihata2013} 
originating from small neutron separation energy or the weak binding of the last neutrons. 
 In particular, 
the pygmy dipole resonance was suggested to influence the r-process nucleosynthesis \cite{Goriely1998}, and
efforts to include the exotic collective excitations has been pursued  in the framework of the CN process models \cite{Goriely1998, Goriely2002, Goriely2004, Litvinova2009, Avdeenkov2011, Daoutidis2012, Xu2014, Martini2016}. However 
the assumption of the statistical compound states may not be appropriate in the case of small neutron separation
energy. Effect of the weak neutron binding is partly taken into account in the direct capture models \cite{Typel2005},
which have however difficulty to include the collective correlations.

The purpose of our study is to formulate  a microscopic model of radiative neutron capture reaction
which is relevant to the r-process. Namely we intend to take into account the 
various nuclear structure effects, which are not included in the conventional direct capture models.
We intend also to describe the whole process of $(n,\gamma)$ reaction on a single quantum many-body theory
so that we can avoid the statistical assumption of the CN process.  
To this end we adopt  the continuum random-phase approximation  (cRPA)
\cite{Shlomo1975, Bertsch1975,Matsuo2001}, which is formulated as a linear response theory 
based on the time-dependent density functional theory (TDDFT) \cite{Nakatsukasa2016}. 
In a previous publication \cite{Matsuo2015}, we have reported
 a prototype formulation which describes
the radiative capture process
$(A -1) + n \rightarrow A^* \rightarrow A_{\rm g.s.} + \gamma$, i.e. a gamma-transition from 
continuum excited states $A^*$, which couples to the neutron scattering state $(A-1) + n$,  to 
the ground state $A_{\rm g.s.}$  of the residual nucleus $A$.
This formulation enables us to describe the collective excitations such as the pygmy and/or giant resonances, or the RPA correlation
in general  in the continuum excited states   $A^*$.  
Note however that the final states of the  actual $(n,\gamma)$ reaction 
are not only the ground state $A_{\rm g.s.}$
but also low-lying excited states $A^{**}$. Therefore, in the present
paper, we extend the formalism so that it can  describe the latter case,  i.e. the
$(n,\gamma)$ reaction $(A -1) + n \rightarrow A^* \rightarrow A^{**} + \gamma$
populating the low-lying excited states $A^{**}$. This is a necessary extension to evaluate the total neutron capture cross section to which
gamma-decays to excited states contribute.
Furthermore we will take into account correlations in the low-lying excited states $A^{**}$, such as surface vibrational modes.

One of the keys of the extension is  given in
a preceding paper \cite{Saito2021}, where we have formulated an extended 
 linear response theory (the continuum random-phase approximation) to describe 
  photo-absorption transitions $A^{**} + \gamma \rightarrow A^*$  from
a low-lying excited state $A^{**} $ to the continuum excited states $A^*$.
 On the basis of this achievement, 
  we formulate in the present paper a framework to calculate the $(n,\gamma$) cross section.
  An important  key for this purpose is  the method of Zangwill and Soven \cite{Zangwill1980},
  which enables us to define the $T$ matrix 
  for photo-absorption followed by particle-emission.
Combining these key formalisms, we describe partial cross sections of $(\gamma,n)$ reaction  
$A^{**} + \gamma \to A^{*} \to n+ (A -1)$ for individual channels of the  
neutron emission  $A^{*} \to n+ (A -1)$.
We then obtain the $(n,\gamma)$ cross section for the inverse process 
$(A -1) + n \rightarrow A^* \rightarrow A^{**} + \gamma$,
using the reciprocity theorem. Details of the formulation is given in Sec. II.

We demonstrate in Sec. III new features of the present theory
by performing  a numerical calculation for radiative neutron capture on $^{139}{\rm Sn}$ with E1 and E2 transitions 
populating low-lying quadrupole and octupole states as well as the ground state in $^{140}{\rm Sn}$. 
 We will show that the presence of the low-lying octupole state brings 
new aspects in the $(n,\gamma)$ reactions originating from strong collectivity of this state. We shall
discuss also that the present cRPA approach describes different kinds of resonance structure emerging in the capture reaction,
including narrow resonances originating from non-collective states as well as the giant resonances and the
single-particle resonances. We draw conclusions in Sec. IV.

\section{Theory}
In this section we formulate a scheme to describe  a radiative neutron capture reaction 
$(A - 1) + n \to A^{*} \to A^{**} + \gamma$ where the final state of the gamma-transition is
 a low-lying excited state  $A^{**}$.  
 We first describe photo-absorption reaction $A^{**} + \gamma \to A^{*}$ assuming that
both initial and final states are described by the continuum random-phase approximation (cRPA) to the
time-dependent density functional theory (TDDFT) (subsections \ref{subsec2A} and \ref{subsec2B}). 
In case the excited state $A^{*}$ is located above the neutron separation energy, $A^{*}$ decays by emitting a neutron $A^{*} \to (A - 1) + n$.
In the framework of the continuum random-phase
approximation the configuration of
the daughter nucleus $(A-1)$ is a one-hole state with respect to the ground state of $A$.
It is then possible to derive partial photo-absorption cross sections for individual channels
of the scattering states  $(A - 1) + n$ by using the Zangwill and Soven method (subsection \ref{subsec2C}).
Finally we obtain the neutron capture cross section for the inverse process $(A - 1) + n \to A^{*} \to A^{**} + \gamma$ using the detailed balance (subsection \ref{subsec2D}).

\subsection{The photo-absorption cross section between RPA excited states}\label{subsec2A}

We express an initial state $A^{**}$ of photo-absorption reaction as $\ket{i{L_{i}M_{i}}}$ with 
excitation energy $E_{i}$, where
 $L_{i}M_{i}$ are the angular momentum numbers.
We also express an excited state $A^{*}$ as $\ket{kLM(E)}$ with  the angular quantum numbers
$LM$. The excitation energy $E$ is shown explicitly  as $A^{*}$ is a state in the continuum spectrum. 
$k$ represents other quantum numbers. Normalization is
$\langle k'L'M'(E') | kLM(E) \rangle =\delta_{k'L'M', kLM} \delta(E-E')$.

The photoabsorption cross section of the transition from $\ket{iL_{i}M_{i}}$ to states with angular momentum $L$ and
energy $E = E_{i} + E_{\gamma}$ is given generally by \cite{Ring1980,Bertulani2004,Thompson2009}
\begin{align}
\sigma^{\lambda}_{iL_{i} + \gamma \to L}(E_{\gamma}) &= f_\lambda(E_{\gamma}) \sum_{k} B(M_{\lambda}, iL_{i} \to kL(E))
=\frac{f_\lambda(E_{\gamma}) }{2L_{i} + 1} S(M_{\lambda}; iL_{i}, L; E) 
\end{align}
for electromagnetic multipole $\hat{M}_{\lambda \mu}$ transition with photon energy $E_{\gamma}$ in terms of
 the reduced matrix element
\begin{align}
B(M_{\lambda}, iL_{i} \to kL(E)) &= \frac{1}{2L_{i} + 1} |\bra{kL(E)} |\hat{M}_{\lambda}| \ket{iL_{i}}|^{2},  
\end{align}
or the strength function
\begin{align}
S(M_{\lambda}; iL_{i}, L; E) &= \sum_{kMM_{i}\mu} |\bra{kLM(E)} \hat{M}_{\lambda \mu} \ket{iL_{i}M_{i}}|^{2}  \notag \\
&= \sum_{k} |\bra{kL(E)} |\hat{M}_{\lambda}| \ket{iL_{i}}|^{2} ,
\end{align}
and a kinematical factor
\begin{align}
f_{\lambda}(E_{\gamma}) &= \frac{(2 \pi)^{3} (\lambda + 1) e^{2}}{\lambda [(2 \lambda + 1) !!]^{2}} \left( \frac{E_{\gamma}}{\hbar c} \right)^{2 \lambda - 1}.
\end{align}
We assume that both excited states $\ket{iL_{i}M_{i}}$ and $\ket{kLM(E)}$ are those
which can be described by cRPA based on the TDDFT. In other words, the model space is spanned
by all the particle-hole configurations including scattering single-particle states.
In the present work we assume that a nucleus $A$ has  closed-shell configurations (or sub-shell-closed configurations) for both neutrons and protons
where the pair correlation can be neglected. Inclusion of the pairing is feasible with use of the method in Ref.\cite{Matsuo2015}, but we leave it for future
publications.

As shown in Ref.\cite{Saito2021},  the strength function
$S(M_{\lambda}; iL_{i}, L; E)$ for transitions from an excited state $iL_{i}$ 
is rewritten as another  strength function 
\begin{align}
 S(F_{L}; g, L; E) \equiv \sum_{k} |\bra{kL(E)} |\hat{F}_{L}| \ket{0^{+}_{\rm g}}|^{2}  = S(M_{\lambda}; iL_{i}, L; E)   
\label{strength_function_for_F}
\end{align}
for an operator
\begin{align}
	\hat{F}_{L M} \equiv \sum_{\mu M_{i}} \langle \lambda \mu L_{i} M_{i} | L M \rangle [\hat{M}_{\lambda \mu}, \hat{O}^{\dag}_{iL_{i}M_{i}}],
\end{align}
defined by a commutator 
between $\hat{M}_{\lambda \mu}$ and the RPA creation operator 
 $\hat{O}^{\dag}_{iL_{i}M_{i}}$ of the low-lying excited state $\ket{iL_{i}M_{i}}=\hat{O}^{\dag}_{iL_{i}M_{i}}\ket{0^{+}_{\rm g}}$. Note that
 the matrix elements in the strength function $S(F_{L};g,L;E)$ are those between the ground state $\ket{0^{+}_{\rm g}}$
 and the RPA excited states $\ket{kLM(E)}$ for the operator $\hat{F}_{L M}$. 
   Since the operator
 $\hat{F}_{LM}$ is a one-body, though non-local, operator,  it is possible to calculate the
 strength function $S(F_{L};g,L;E)$  using an extended linear response formulation of the cRPA\cite{Saito2021}.

\subsection{Linear response formalism}\label{subsec2B}

Here we recapitulate briefly an essence of the formulation\cite{Saito2021} by omitting the angular momentum
algebra.

The non-local one-body operator $\hat{F} = [\hat{M}, \hat{O}^{\dag}_{i}]$ is expressed as
\begin{align}
	\hat{F} = \iint dx dy F(x, y) \psi^{\dag}(x) \psi(y),
\end{align}
where $\psi^{\dag}(x)$ and $\psi(y)$ are the creation and annihilation operators of the nucleon, and
 $F(x,y)$ is the coordinate representation of the matrix element.
Here $x$ ( and $y$) is shorthand notation of the space coordinate $\vec{r}$ and the spin 
variable $\sigma = \uparrow, \downarrow$, i.e. $x = \vec{r} \sigma$, $\int dx = \sum_{\sigma} \int d \vec{r}$.
Using the linear response $\delta \rho(x, y, \omega)$ of density matrix $\rho(x,y) = \langle \hat{\rho}(x,y) \rangle$, $\hat{\rho}(x,y) = \psi^{\dag}(y) \psi(x)$ for the perturbation $\hat{F}$, we can calculate the strength function $S(F; \hbar \omega) = \sum_{k} |\bra{k(E)} \hat{F} \ket{0}|^{2}$ \ \ $ (E = \hbar \omega)$ as
\begin{align}
	S(F; \hbar \omega)  =  - \frac{1}{\pi} {\rm Im} \iint dxdy \, F^{*}(x, y) \delta \rho(x, y, \omega).
\end{align}

In TDDFT scheme
the density matrix response $\delta \rho(x, y, \omega)$ obeys the linear response equation
\begin{align}
	\delta \rho(x, y, \omega) = \iint dx^{'} dy^{'} R_{0}(x, y; y^{'}, x^{'}; \omega) (F(x^{'}, y^{'}) 
	+v_{\rm ind}(x^{'},  \omega)\delta(x^{'} - y^{'})).
\label{linear_response_equation_delta_rho}
\end{align}
Here $v_{\rm ind}(x, \omega)$ is a time-dependent part of the Hartree-Fock (or Kohn-Sham) potential originating from the density response,
which is called the induced field.
Assuming that the HF (KS)  mean-field $U[\rho](x)$ is local, it is given by
\begin{align}
v_{\rm ind}(x, \omega)=\frac{\delta U(x)}{\delta \rho}\delta\rho(x,\omega)
\end{align}
with $\delta\rho(x,\omega)\equiv\delta\rho(x,x,\omega)$. The  function
$R_{0}$ is an unperturbed response function for the density matrix, and given by
\begin{align}
\label{R_0_spectrum}
	R_{0}(x, y; y^{'}, x^{'}; \omega) \equiv \sum_{ph} \left\{ \frac{\bra{0} \hat{\rho}(x, y) \ket{ph} \bra{ph} \hat{\rho}(y^{'}, x^{'}) \ket{0}}{\hbar \omega - (\epsilon_{p} - \epsilon_{h}) + i \eta} - \frac{\bra{0} \hat{\rho}(y^{'}, x^{'}) \ket{ph} \bra{ph} \hat{\rho}(x, y) \ket{0}}{\hbar \omega + (\epsilon_{p} - \epsilon_{h}) + i \eta} \right\}
\end{align}
in the spectral representation in terms of the single-particle eigenstates in the static HF (KS) potential $U[\rho_0](x)$.
$\ket{ph} = a^{\dag}_{p} a_{h} \ket{0}$ is one-particle-one-hole configuration, $\epsilon_{p}$ and $\epsilon_{h}$ are single-particle energies of the particle and hole orbits.
$\eta$ is a positive infinitesimal constant.

The response function $R_{0}$ is expressed also as 
\begin{align}
\label{R_0_Green}
	R_{0}(x, y; y^{'}, x^{'}; \omega) = \sum_{h} &\left\{ \phi^{*}_{h}(y)G_{0}(x, x^{'}, \epsilon_{h} + \hbar  \omega + i\eta)\phi_{h}(y^{'}) \right. \notag \\ 
&\left. + \phi^{*}_{h}(x^{'})G_{0}(y^{'}, y, \epsilon_{h} - \hbar \omega - i\eta)\phi_{h}(x) \right\},
\end{align}
where $G_{0}(x, x^{'}, \epsilon)$ is the single-particle Green's function for the static HF (KS) potential $U_0$,
and $\phi_{h}(x)$ is the wave function of the hole orbit $h$. The Green's function satisfies
a proper asymptotic boundary condition for $|x| , |x^{'}| \to \infty$ so that it describes scattering
waves for unbound orbits. 

We refer the readers to Ref.\cite{Saito2021} for equations with the angular momentum algebra.

\subsection{The Zangwill-Soven method for partial photo-absorption cross section}\label{subsec2C}

In the  linear response formalism discussed above, unbound particle states $p$ represent the scattering states.
 It is therefore possible to describe decay of the RPA excited states
with  emission of a neutron (or a proton). 
Following Zangwill and Soven \cite{Zangwill1980},
we shall calculate  partial  $(\gamma,n)$ cross section
for individual decay channels.

We first note that the strength function Eq. (\ref{strength_function_for_F}) is rewritten as
\begin{align}
	S(\hat{F}; \hbar \omega) &= - \frac{1}{\pi} {\rm Im} \iiiint dx dy dx^{'} dy^{'} v^{*}_{\rm scf}(x, y, \omega) R_{0}(x, y ; y^{'}, x^{'} ; \omega) v_{\rm scf}(x^{'}, y^{'}, \omega) \notag \\
&= \sum_{ph} \left| \bra{ph} \hat{V}_{\rm scf}(\hat{F}; \omega) \ket{0} \right|^{2} \delta(\hbar \omega - (\epsilon_{p} - \epsilon_{h})).
\label{strength_function_for_Vscf}
\end{align}
Here
$\hat{V}_{\rm scf}(\hat{F}; \omega)$ is a one-body field, called the self-consistent field \cite{Zangwill1980,Nakatsukasa2001}, defined as a sum of induced field $\hat{V}_{\rm ind}$ and the perturbing field $\hat{F}$:
\begin{align}
\label{def_Vscf}
	\hat{V}_{\rm scf}(\hat{F}; \omega) &\equiv \hat{F} + \hat{V}_{\rm ind}(\hat{F}; \omega) 
= \iint dx dy \, v_{\rm scf}(x, y, \omega) \hat{\rho}(y, x) , \\
v_{\rm scf}(x, y, \omega) &\equiv F(x, y) + \frac{\delta U(x)}{\delta \rho} \delta\rho(x, \omega)\delta(x-y).
\end{align}

\begin{figure}[H]
	\centering
	\includegraphics[width=0.7\columnwidth]{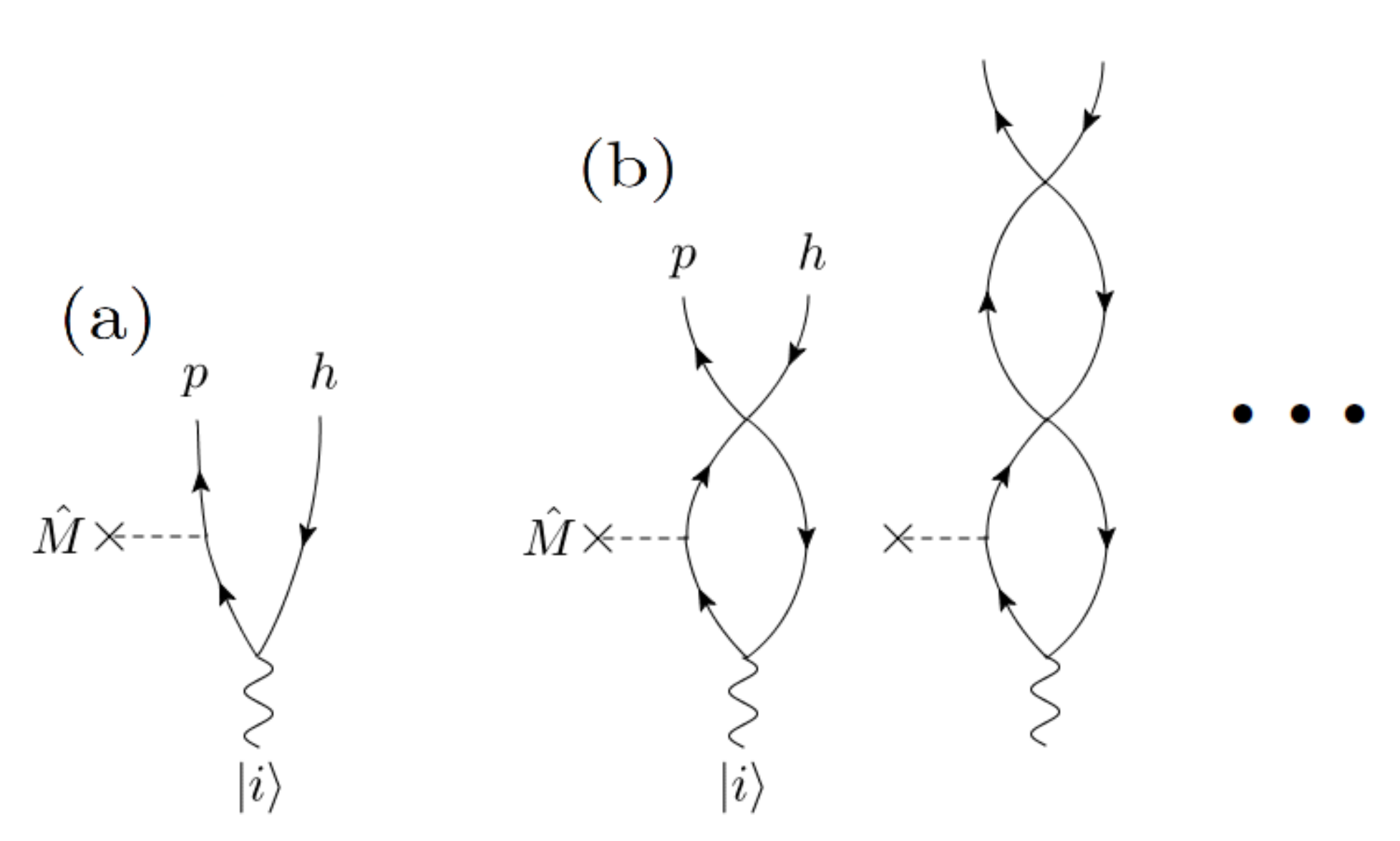}
	\caption{
The diagrams representing the matrix element $\bra{ph} \hat{V}_{\rm scf}(\hat{F}; \omega) \ket{0}$.
(a) and (b) correspond to the first and second terms of r.h.s. of Eq.(\ref{matel_Vscf}), respectively. Note that
the electromagnetic operator $\hat{M}$ acts also on the hole line.}
	\label{Fdiagram}
\end{figure}

Note that matrix element  $\bra{ph} \hat{V}_{\rm scf}(\hat{F}; \omega) \ket{0}$ is a $T$ matrix for the
$(\gamma,n)$ reaction under consideration. This is seen in
\begin{align}
\label{matel_Vscf}
	\bra{ph} \hat{V}_{\rm scf}(\hat{F}; \omega) \ket{0} = \bra{ph} \hat{F} \ket{0} + \int dx \bra{ph} \frac{\delta U(x)}{\delta \rho}\hat{\rho}(x) \ket{0} \delta \rho(x, \omega).
\end{align}
The first term $\bra{ph} \hat{F} \ket{0}$ corresponds to the diagram representation shown in Fig. \ref{Fdiagram}(a).
The density response $\delta \rho(x, \omega)$ in the second term can be expanded in infinite series
\begin{align}
\label{infinite_series_delta_rho}
	\delta \rho = R_{0}F +  R_{0}\frac{\delta U}{\delta \rho} \delta \rho = R_{0} F + R_{0} \frac{\delta U}{\delta \rho} R_{0} F + R_{0} \frac{\delta U}{\delta \rho} R_{0} \frac{\delta U}{\delta \rho} R_{0} F + \cdots,
\end{align}
representing symbolically  the linear response equation (\ref{linear_response_equation_delta_rho}).
We notice that this infinite series correspond to  ring diagrams, and the matrix element 
$\bra{ph} \hat{V}_{\rm scf}(\hat{F}; \omega) \ket{0}$ is represented by the diagrams shown in Fig. \ref{Fdiagram}.

These diagrams indicate that 
the matrix element $\bra{ph} \hat{V}_{\rm scf}(\hat{F}; \omega) \ket{0}$ represents
the transition amplitude for the electromagnetic operator $\hat{M}$
between the low-lying excited state $\ket{i}$ and a RPA excited state (represented
by series of the ring diagrams) which is connected to a specific particle-hole
configuration $\ket{ph}$. 
 In the case when the particle state $p$ is an unbound single-particle state, it represents
 the transition-matrix, $T$ matrix, of the $(\gamma,n)$ reaction where the final state 
 consists of
 a unbound neutron (specified by the particle configuration $p$) and the residual nucleus
 $(A-1)$ with one-hole configuration $a_h\ket{0}$. We call it  the RPA $T$ matrix.
We give more detailed discussion in Appendix A.
 
Consequently each term in the r.h.s. of Eq. (\ref{strength_function_for_Vscf}) is related to 
the partial photoabsorption cross section for a $(\gamma,n)$ process,   $A$ (state $i$)  $+\gamma \to 
(A-1)$ (one-hole state $a_h\ket{0}$) $ + n$  (single-particle state $p$):
\begin{align}
\sigma^{\lambda}_{i + \gamma \to ph}(E_{\gamma}) = f_{\lambda}(E_{\gamma}) \left| \bra{ph} \hat{V}_{\rm scf}(\hat{F}; \omega) \ket{0} \right|^{2} \delta(\hbar \omega - (\epsilon_{p} - \epsilon_{h})).
\end{align}
with $E_\gamma = \hbar\omega - E_i$.

The photo-absorption cross section can written also 
 in terms of the single-particle Green's function.
Inserting Eq.(\ref{R_0_Green}) to
Eq.(\ref{strength_function_for_Vscf}), we have
the partial photo-absorption cross section:
\begin{align}
	\sum_{p(\epsilon_p=\epsilon) } \sigma^{\lambda}_{i + \gamma \to ph}(E_{\gamma}) &= - \frac{f_\lambda(E_{\gamma})}{\pi} {\rm Im} \iiiint dx dy dx^{'} dy^{'} \phi^{*}_{h}(y) v^{*}_{\rm scf}(x, y, \omega) \notag \\
&\qquad \qquad \times G_{0c}(x, x^{'}, \hbar \omega + \epsilon_{h} + i \eta) v_{\rm scf}(x^{'}, y^{'}, \omega) \phi_{h}(y^{'}),
\label{neutron_emi_cross_section}
\end{align}
for final states consisting of an escaping neutron with energy $\epsilon = \hbar \omega+ \epsilon_{h}=E_\gamma + E_i + \epsilon_{h}$ and a one-hole state $a_{h} \ket{0}$ of the residual nucleus $(A-1)$.
Here $G_{0c}$ is the  single-particle Green's function in which a contribution of bound orbits is removed:
\begin{align}
\label{G_0c}
	G_{0c}(x, x^{'}, \epsilon) \equiv G_{0}(x, x^{'}, \epsilon) - \sum_{i(\epsilon_{i} < 0)} \frac{\phi_{i}(x) \phi^{*}_{i}(x^{'})}{\epsilon - \epsilon_{i}}.
\end{align}

\subsection{Concrete expressions for $(\gamma,n)$ and $(n,\gamma)$ cross sections}\label{subsec2D}

In actual application we consider the angular momentum algebra with the polar-coordinate representation.
For example $v_{\rm scf}(x, y, \omega)$ is represented as 
\begin{align}
v^{\rm scf}_{L M}(x, y, \omega) &= \sum_{ljm, l^{'}j^{'}m^{'}} Y_{l^{'}j^{'}m^{'}}(\hat{x}) \frac{1}{\sqrt{2j^{'} + 1}} \langle j m L M| j^{'} m^{'} \rangle \frac{v^{{\rm scf}}_{L, l^{'}j^{'}, lj}(r_{x}, r_{y})}{r_{x} r_{y}} Y^{*}_{ljm}(\hat{y}).
\end{align}
Here $Y_{ljm}$ is the spin spherical harmonics.
We specify the escaping neutron with energy $\epsilon$ and the partial wave quantum numbers $lj$,
and express the final state 
$\ket{ph}$ as $\ket{[\epsilon l_{p}j_{p}, n_{h}l_{h}j_{h}]_{LM}} = \sum_{m_{p}m_{h}} \braket{j_{p}m_{p}j_{h}m_{h} | LM} a^{\dag} _{\epsilon l_{p}j_{p}m_{p}} a_{\widetilde{{n_{h}l_{h}j_{h}m_{h}}}} \ket{0}$ with $L$ being the
total angular momentum, which is identical to that of the RPA excited state $A^{*}$.
Accordingly we obtain the expression for the partial photo-absorption cross section
for the specific channel of neutron decay as
\begin{align}
	\sigma^{\lambda}_{iL_{i} + \gamma \to [\epsilon l_{p}j_{p}h]_L}(E_{\gamma}) &= f_{\lambda}(E_{\gamma}) \frac{1}{2L_{i} + 1} \left| \langle [\epsilon l_{p}j_{p}h]_{L} || \hat{V}_{\rm scf}(\hat{F}_{L}; \omega) || 0 \rangle \right|^{2} \notag \\
&= - \frac{f_\lambda(E_{\gamma})}{\pi(2L_{i} + 1)} \notag \\
&\quad \times {\rm Im} \iiiint dr_{x} dr_{y} dr_{x^{'}} dr_{y^{'}} \phi^{*}_{n_{h}l_{h}j_{h}}(r_{y}) v^{{\rm scf} *}_{L, l_{p}j_{p}, l_{h}j_{h}}(r_{x}, r_{y}, \omega) \notag \\
&\quad \times G_{0c, l_{p}j_{p}}(r_{x}, r_{x^{'}}, \epsilon_{h} + \hbar\omega+ i \eta) v^{\rm scf}_{L, l_{p}j_{p}, l_{h}j_{h}}(r_{x^{'}}, r_{y^{'}}, \omega) \phi_{n_{h}l_{h}j_{h}}(r_{y^{'}}).
\label{neutron_emi_cross_section_radial}
\end{align}
Note that 
$\hbar\omega= E_\gamma+E_{i}$ is the excitation energy of the state $A^{*}$ of the nucleus
$A$. Expression of $v^{\rm scf}_{L, l_{p}j_{p}, l_{h}j_{h}}(r_{x^{'}}, r_{y^{'}}, \omega)$ and a closed
form expression of the cross section are given in Appendix B.

Finally, we use the detailed balance to obtain the radiative neutron capture cross section for the reaction
$(A - 1) + n \to A^{*} \to A^{**} + \gamma$ :  
\begin{align}
	\sigma^\lambda_{[\epsilon l_{p}j_{p}h]_{L} \to iL_{i} + \gamma}(\epsilon) &= \frac{2L_{i} + 1}{2j_{h} + 1} \frac{E^{2}_{\gamma}}{2mc^{2} \epsilon} \sigma^{\lambda}_{iL_{i} + \gamma\to [\epsilon l_{p}j_{p}h]_L}(E_{\gamma}).
\end{align}
for an incident neutron with partial wave $l_pj_p$ and energy $\epsilon$ impinging 
on the $(A-1)$ nucleus with a one-hole  configuration $h$, and gamma transition to the excited state 
$\ket{iL_{i}M_{i}}$.
 The total cross section is obtained by
summing all the contributions of the partial waves $l_p j_p$ and the total angular momentum $L$
of the system.  

\section{Numerical examples}
\subsection{Setting}
We shall apply the theory to radiative neutron capture reaction on a neutron-rich nucleus $^{139}{\rm Sn}$, which populates
the ground and low-lying excited states of $^{140}{\rm Sn}$ after gamma transition. 
We have chosen these isotopes since  the neutron separation energy of $^{140}{\rm Sn}$
 is estimated to be $\sim 3$ MeV \cite{Massexpl} and these isotopes
are relevant to the r-process. Another reason is that $^{140}{\rm Sn}$ has a subshell closure at the $2f_{7/2}$ 
orbit for neutrons,
and the pairing correlation neglected in the present formulation is expected to be weak.

Numerical calculations are performed with the same setting as in our previous paper \cite{Saito2021}. 
We use a Woods-Saxon potential in place of the static self-consistent field $U_{0}$ and a Skyrme-type contact interaction as the residual two-body force, given by 
\begin{align}
v_{ph}(\mathbf{r}, \mathbf{r}^{'}) = \left\{t_{0}(1 + x_{0} P_{\sigma}) + \frac{t_{3}}{12}(1 + x_{3} P_{\sigma}) \rho(r)\right\} \delta(\mathbf{r} - \mathbf{r}^{'}).
\label{v_ph}
\end{align}
The details including the parameters are the same as those in \cite{Saito2021}. 
We describe the single-particle wave function 
 by solving the radial Schr\"{o}dinger equation with the Runge-Kutta method 
 up to a maximal radius $R_{\rm max} = 20$ fm (with interval $\Delta r = 0.2$ fm). At $r = R_{\rm max}$ 
 the single-particle wave function is connected to the asymptotic wave, i.e. the Hankel function 
 with an appropriate (complex) wave number. 
 The small real constant $\eta$ in the response equation is set to $\eta = 1.0 \times 10^{-5}$ or $0.1$ MeV.

In the present analysis, we intend to demonstrate effects  of the RPA correlations on the $(n,\gamma)$ reaction, in particular,
roles of the collectivity which may exist both in the scattering state  of $^{139}{\rm Sn}+n $ and the final states of
$^{140}{\rm Sn} $.
We focus on  the low-lying quadrupole and octupole vibrational states as well as 
the dipole and quadrupole giant resonances. Other types of the particle-hole excitations such as the soft dipole mode and non-collective excitations are also discussed. We take into account the RPA correlation for the states with natural spin-parity $L^{\pi}=1^-,2^+$ and $3^-$ whereas we neglect it for other spin-parities.

Table \ref{spe_140Sn_2} shows the Woods-Saxon single-particle orbits. 
The highest occupied neutron orbit in $^{140}{\rm Sn}$ is  $2f_{7/2}$ with $\epsilon_{2f_{7/2}} = -2.59$ MeV (the Fermi energy). The configuration of
the ground state of $^{139}{\rm Sn}$ is assumed to be a configuration with neutron one-hole in $2f_{7/2}$ with spin-parity
$j_{h}^{\pi}=7/2^{-}$. We denote it  $^{139}{\rm Sn}({\small 7/2}^{-})$ in the following.
The RPA calculation for $^{140}{\rm Sn}$ brings about several bound excited states 
below the neutron separation energy $S_{1n}=2.59$ MeV (see Fig. 2 of Ref. \cite{Saito2021}).
Among them we adopt the lowest-lying excited states, i.e.  $2^{+}_{1,2}$ and $3^{-}_{1}$ states, as the final states $\ket{i L_{i}}$ of the 
$^{139}{\rm Sn}(n,\gamma)^{140}{\rm Sn}$ reaction. The excitation energy of these states are
$E_{2^{+}_{1}} = 0.888$ MeV,$ \, E_{2^{+}_{2}} = 1.093$ MeV,$ \, E_{3^{-}_{1}} = 1.768$ MeV. As seen in the
RPA forward amplitudes $X_{ph}$ of these states (Tables \ref{tableXamp2} and \ref{tableXamp3}), 
the $3^{-}_{1}$ state exhibits a moderately strong collectivity, i.e. it contains many particle-hole configurations
both in neutrons and protons, typical of the surface vibrational octupole state. The  $2^{+}_{1}$ 
and $2^{+}_{2}$ states consist mostly of two neutron particle-hole configurations,
$(1h_{9/2})(2f_{7/2})^{-1}$ and $(3p_{3/2})(2f_{7/2})^{-1}$ while the correlation causes strong
mixing among them. 

For the gamma transition, we mostly discuss the E1 transition, but we also mention briefly the E2 case.
We take into account all possible angular and spin quantum numbers of  the
total system  (both for the scattering state  of $^{139}{\rm Sn}+n $ and the final states of
$^{140}{\rm Sn} $) as long as they are allowed 
by the angular momentum coupling of
the adopted final states ($L_i^{\pi}=0_{\rm g.s.}^{+},  2^{+}_{1,2}$ and $3^{-}_{1}$) and the multipolarity
($\lambda=1,2$) of the gamma ray. Namely, we describe
$^{139}{\rm Sn}({\small 7/2}^{-}) + n \to ^{140}{\rm Sn}(1^{-}) \to ^{140}{\rm Sn(g.s.)} + \gamma$,$^{139}{\rm Sn}({\small 7/2}^{-}) + n \to ^{140}{\rm Sn}(1^{-},2^{-},3^{-}) \to ^{140}{\rm Sn(2^{+}_{1,2})} + \gamma$,$^{139}{\rm Sn}({\small 7/2}^{-}) +n \to ^{140}{\rm Sn}(2^{+},3^{+},4^{+}) \to ^{140}{\rm Sn(3^{-}_{1})} + \gamma$ for the E1 transition
while $^{139}{\rm Sn}({\small 7/2}^{-}) + n \to ^{140}{\rm Sn}(2^{+}) \to ^{140}{\rm Sn(g.s.)} + \gamma$,$^{139}{\rm Sn}({\small 7/2}^{-}) + n \to ^{140}{\rm Sn}(0^{+},1^{+},2^{+},3^{+},4^{+}) \to ^{140}{\rm Sn(2^{+}_{1,2})} + \gamma$,$^{139}{\rm Sn}({\small 7/2}^{-}) +n \to ^{140}{\rm Sn}(1^{-},2^{-},3^{-},4^{-},5^{-}) \to ^{140}{\rm Sn(3^{-}_{1})} + \gamma$.
for the E2 case.  We include all the partial waves of the incident neutron which are allowed by the coupling to the
the spin-parity $7/2^{-}$ of the target nucleus $^{139}{\rm Sn}$.  Its maximum is $l_p, j_p=8, 17/2$. 

\begin{table}
	\centering
	\caption{
		Single-particle energies of the adopted Woods-Saxon potential for $^{140}{\rm Sn}$. 
		Several orbits around the Fermi energy (indicated by lines) are listed.}
	\label{spe_140Sn_2}
\begin{tabular}[t]{ c c p{5mm} c c } \hline \hline
neutron & $\epsilon \, [{\rm MeV}]$ & & proton &  $\epsilon \, [{\rm MeV}]$ \\ \hline
			$2f_{5/2}$ &   -0.31 && $1h_{11/2}$ &  -11.40\\		
			$3p_{1/2}$ &   -0.81 && $2d_{3/2}$ &  -11.61 \\ 
			$3p_{3/2}$ &   -1.46 && $2d_{5/2}$ &  -14.06\\ 
			$1h_{9/2}$ &   -1.53  && $1g_{7/2}$ &  -15.08\\ \cline{1-2}\cline{4-5}
			$2f_{7/2}$ &   -2.59 && $1g_{9/2}$ &  -19.97\\  
			$1h_{11/2}$ &   -6.64 && $2p_{1/2}$ &  -21.75\\ 
			$3s_{1/2}$ &   -8.65 && $2p_{3/2}$ &  -23.02\\ 
			$2d_{3/2}$ &   -8.65  && $1f_{5/2}$ &  -24.81\\ 
			$2d_{5/2}$ &  -10.40 &&& \\ 
			$1g_{7/2}$ &  -10.96 &&& \\ 
			$1g_{9/2}$ &  -14.64 &&& \\ \hline \hline
\end{tabular}
\end{table}

\begin{table}
	\centering
	\caption{
		The RPA forward amplitudes $X_{ph}$ of the $2^{+}_{1}$ and $2^{+}_{2}$ state. 
		Particle-hole configurations with large amplitude $|X_{ph}| > 0.1$ are listed.
		The RPA backward and forward amplitudes $X_{ph}$ and $Y_{ph}$ are calculated using a method of Ref.\cite{Shimoyama2013}.}
    	\begin{tabular}[t]{ c c c } \hline \hline
		neutron config. & $X_{ph}^{2^{+}_{1}}$ & $X_{ph}^{2^{+}_{2}}$ \\ \hline 
		$(1h_{9/2})(2f_{7/2})^{-1}$ & -0.601 & 0.791 \\ 
		$(3p_{3/2})(2f_{7/2})^{-1}$ & 0.789 & 0.600  \\ \hline \hline
   	\end{tabular}
	\label{tableXamp2}
\end{table}

\begin{table}
	\centering
	\caption{
		The RPA forward amplitudes $X_{ph}$ of the $3^{-}_{1}$ state. 
		Particle-hole configurations with large amplitude $|X_{ph}| > 0.1$ are listed.
		The neutron single-particle orbit $1i_{13/2}$ is a resonance in the continuum.}
    		\begin{tabular}[t]{ c c p{5mm} c c} \hline \hline
		      	neutron config. &  $X^{3^{-}_{1}}_{ph}$ & &  proton config. &  $X^{3^{-}_{1}}_{ph}$ \\ \hline 
			$(1i_{13/2})(2f_{7/2})^{-1}$ & 0.831 && $(1h_{11/2})(1g_{9/2})^{-1}$ & -0.285  \\ 
			$(1i_{13/2})(1h_{11/2})^{-1}$ & 0.354 && $(1g_{7/2})(2p_{1/2})^{-1}$ & 0.203  \\ 
			$(1h_{9/2})(2d_{3/2})^{-1}$ & -0.299 && $(2d_{5/2})(2p_{1/2})^{-1}$ & 0.176  \\ 
			$(1h_{9/2})(1g_{7/2})^{-1}$ & -0.189 && $(2d_{5/2})(2p_{3/2})^{-1}$ & 0.135  \\
			$(2f_{5/2})(3s_{1/2})^{-1}$ & 0.134 && $(1j_{15/2})(1g_{9/2})^{-1}$ & 0.129  \\ 
			$(2g_{9/2})(2f_{7/2})^{-1}$ & 0.133 && $(2f_{7/2})(1g_{9/2})^{-1}$ & -0.129  \\ 
			$(2f_{5/2})(2d_{3/2})^{-1}$ & -0.126 && $(2d_{3/2})(2p_{3/2})^{-1}$ & -0.120  \\ 
			$(3p_{3/2})(2d_{3/2})^{-1}$ & -0.112 && $(3p_{3/2})(1g_{9/2})^{-1}$ & -0.103  \\ 
			$(2j_{15/2})(1g_{9/2})^{-1}$ & 0.108 && $(1g_{7/2})(1f_{5/2})^{-1}$ & 0.102  \\ 
			$(2f_{5/2})(1g_{7/2})^{-1}$ & -0.101 && & \\ \hline \hline
   		 \end{tabular}
	\label{tableXamp3}
\end{table}   

\subsection{Low energy $(n,\gamma)$ cross section with E1 transition}
\begin{figure}
\centering
\begin{minipage}{\columnwidth}
\includegraphics[width=0.52\columnwidth]{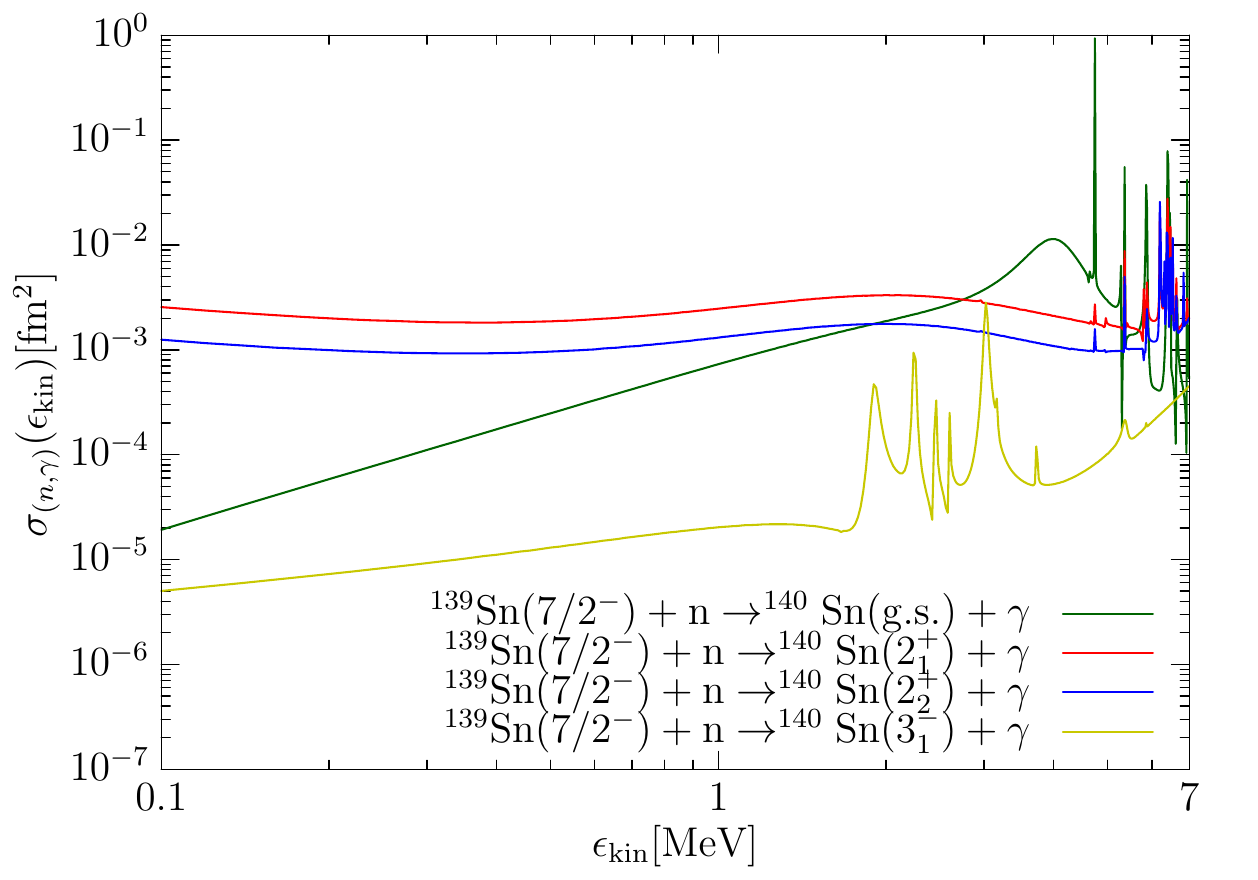}
\end{minipage} 
\caption{
The calculated $(n,\gamma)$ cross sections for $^{139}{\rm Sn}({\small 7/2}^{-})+ n \to ^{140}{\rm Sn} + \gamma$
with  E1 transitions populating different low-lying states in $^{140}{\rm Sn}$; the ground state (plotted with a green curve),
the low-lying $2^{+}_{1}$ and $2^{+}_{2}$ states (red and blue curves respectively), and
the octupole vibrational state $3^{-}_{1}$ (yellow curve). 
The horizontal axis is kinetic energy $\epsilon_{\rm kin}$ of the incident neutron.
The smoothing parameter  is $\eta=10^{-5}$ MeV.}
 \label{E1_ncap_total}
\end{figure}

Figure \ref{E1_ncap_total} shows the cross sections of neutron capture with E1 
for  low neutron kinetic energy relevant to the r-process,
calculated separately for each final state.
A characteristic feature is that the cross sections exhibit significant differences for different final states.
In particular most dominant transition at low energy is the ones to the $2_1^{+}$ and
$2_2^{+}$ states rather than to the ground state. Another noticeable feature is that the transition to the
octupole vibrational state $3^{-}_{1}$  is present although the absolute magnitude is small.  
It is noted that
there is no negative parity one-particle one-hole ($1p1h$) configuration with energy smaller  the neutron separation energy $2.59$ MeV.
The collective $3^{-}_{1}$ is exceptional as this state emerges at low energy due to the RPA correlation.
The third observation is that the cross sections 
exhibit many resonance-like behaviors
above $~2$ MeV whereas they are 
smooth  below $~2$ MeV.  In the following we shall discuss these features in more details.

\subsubsection{Transitions to the ground state}\label{discuss_gs}

\begin{figure}
\centering
\begin{minipage}{\columnwidth}
\includegraphics[width=0.52\columnwidth]{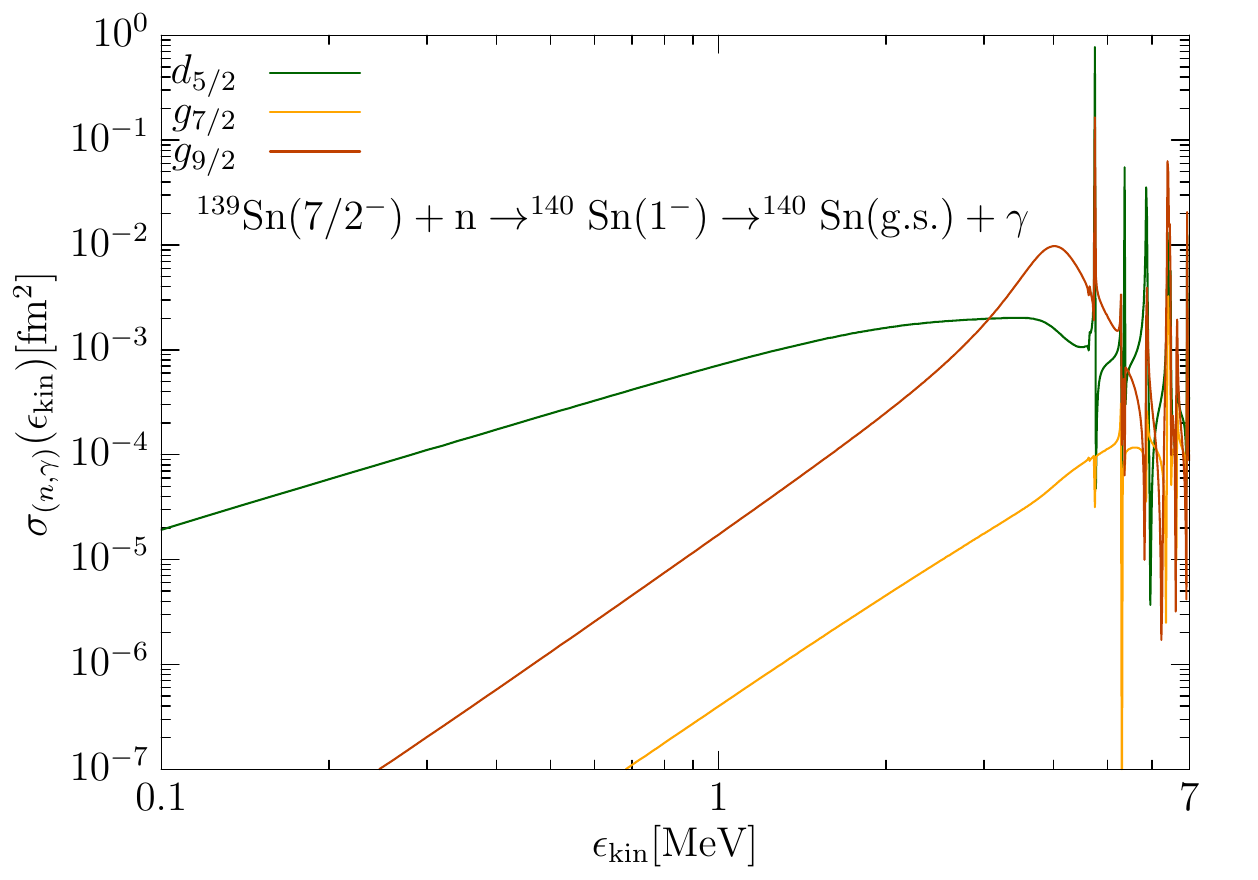}
\end{minipage} 
\caption{
The calculated partial $(n,\gamma)$ cross sections for
$^{139}{\rm Sn}({\small 7/2}^{-}) + n \to ^{140}{\rm Sn}(1^{-}) \to ^{140}{\rm Sn(g.s.)} + \gamma$ for E1 transitions,
plotted separately for different partial waves of the incident neutron; $d_{5/2}$ (green curve), $g_{7/2}$ (orange),
and $g_{9/2}$ (brown). The horizontal axis is the neutron kinetic energy $\epsilon_{\rm kin}$.
The smoothing parameter  is $\eta=10^{-5}$ MeV.
}
\label{E1_ncap_decay_to_gs}
\end{figure}

Let us discuss the $(n,\gamma)$ reaction where the gamma transition  populates directly the ground state of $^{140} {\rm Sn}$, whose
cross section is the green curve in Fig. \ref{E1_ncap_total}. 
In order to analyze the structures seen in Fig. \ref{E1_ncap_total}, we decompose the cross section
with respect to the partial waves of the incident neutron. The decomposed cross sections are shown in 
Fig. \ref{E1_ncap_decay_to_gs}. In the present case the relevant spin-parity of the total system $ n + ^{139} {\rm Sn}$ or that of excited continuum state of $^{140} {\rm Sn}$
is  $1^{-}$. Thus the partial waves of the incident neutron are limited to $d_{5/2}$ and $g_{7/2,9/2}$ 
due to the angular momentum coupling to 
 the ground state  $^{139}{\rm Sn}({\small 7/2}^{-}) $.

The cross section at low energy $\le 1$ MeV is dominated by
the $d_{5/2}$-wave since it has the lowest orbital angular momentum among the available partial waves. 
The  $s$-wave capture is forbidden due to the angular coupling rule.
The low energy
behavior of the partial cross sections exhibits a scaling  $\propto \epsilon^{l_p-1/2}$, which points to 
direct transitions from non-resonant partial waves  to the ground state of $^{140} {\rm Sn}$.
It is seen that a rather broad peak around $\epsilon_{\rm kin}=4$ MeV is that in the $g_{9/2}$-wave. 
It is the E1 transition from the single-particle $g_{9/2}$ resonance
to the bound $2f_{7/2}$ orbit which is vacant in $^{139}{\rm Sn}$. These features
reflect the single-particle nature of the soft dipole excitation of $^{140}{\rm Sn}$.

\begin{figure}
\centering
\begin{minipage}{\columnwidth}
\includegraphics[width=0.3\columnwidth]{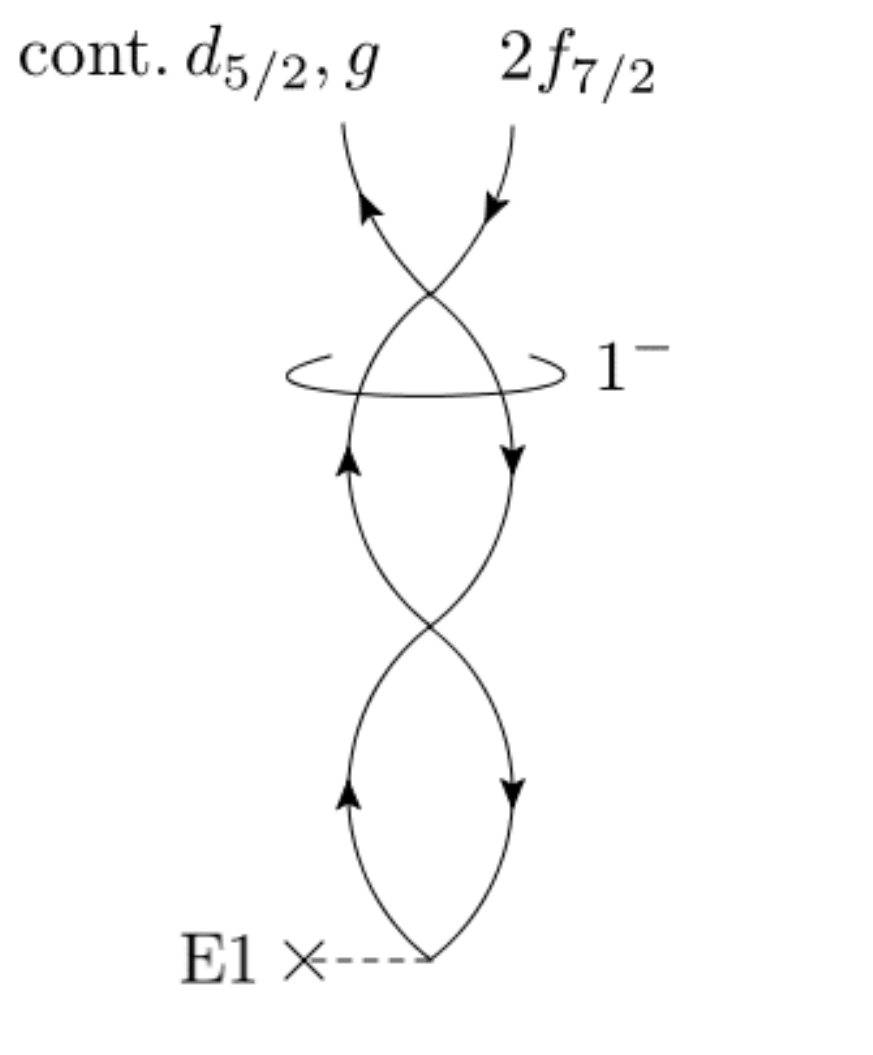}
\end{minipage} 
\caption{The diagrams representing $(n,\gamma)$ reaction of $^{139}{\rm Sn}({\small 7/2}^{-}) + n \to ^{140}{\rm Sn(1^{-})} \to ^{140}{\rm Sn(g.s.) + \gamma}$. }
\label{diagram_gs}
\end{figure}

Several sharp peaks in the energy range   
$\epsilon_{\rm kin}  > 4.5 $ MeV are
however cannot be of the single-particle origin. Indeed they originates from non-collective states whose main components are one-particle-one-hole ($1p1h$) configurations
$\nu[(3p_{3/2})(3s_{1/2})^{-1}]$, $\nu[(3p_{1/2})(3s_{1/2})^{-1}]$, $\nu[(3p_{3/2})(2d_{3/2})^{-1}]$, $\nu[(3p_{1/2})(2d_{3/2})^{-1}]$, $\nu[(3p_{3/2})(2d_{5/2})^{-1}]$, and $\nu[(3p_{1/2})(2d_{5/2})^{-1}]$. Although the particle and hole orbits of these configurations are both bound orbits (see Table \ref{spe_140Sn_2}),  they couple to the scattering states
of $ n + ^{139}{\rm Sn}$ via the residual interaction as is represented in Fig.\ref{diagram_gs}, and they form
narrow resonances. 
Note that the non-collective states with proton $1p1h$ configurations also appear as narrow resonances.

\begin{figure}
\centering
\begin{minipage}{\columnwidth}
\includegraphics[width=0.52\columnwidth]{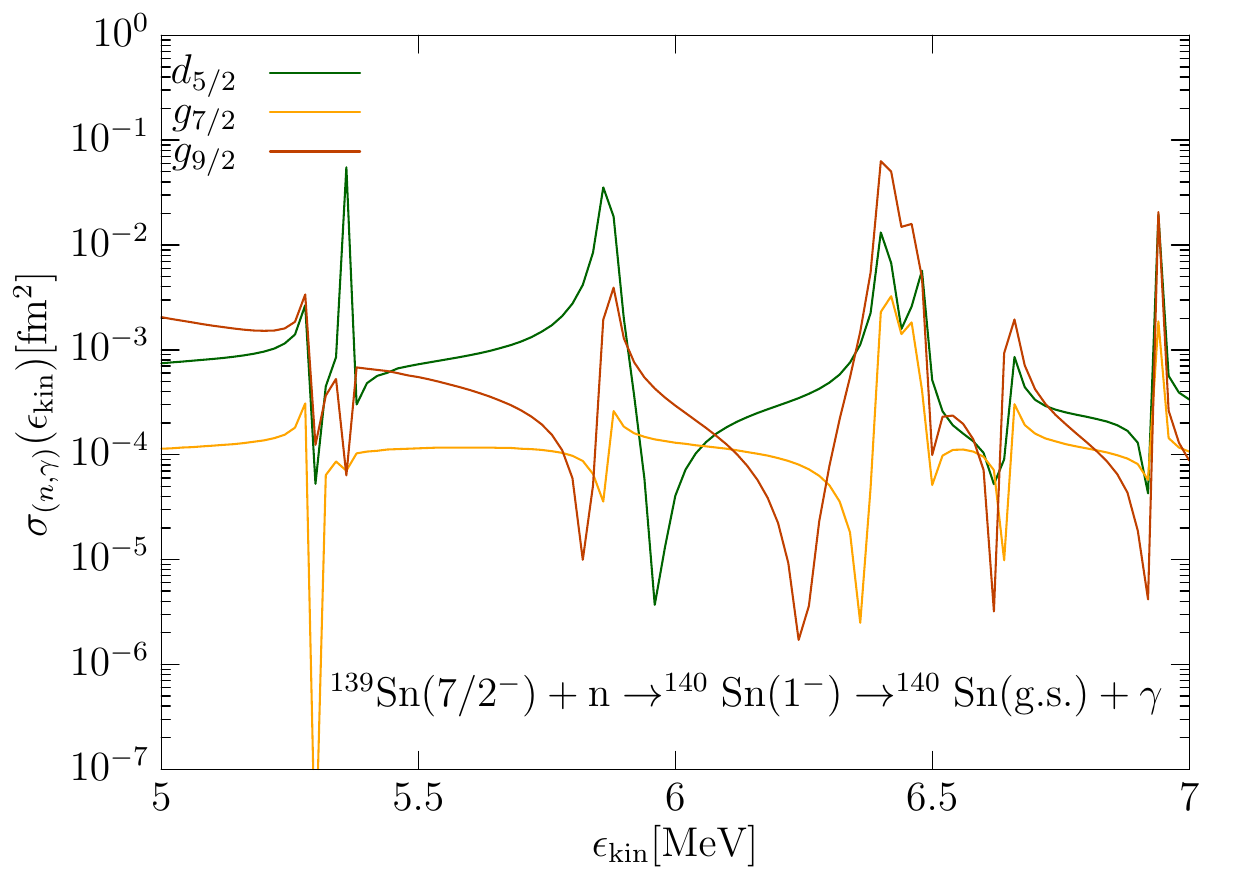}
\end{minipage} 
\caption{
The same as Fig. \ref{E1_ncap_decay_to_gs}, but a magnification in the range of $\epsilon_{\rm kin} = 5.0 - 7.0$ MeV.
}
\label{E1_ncap_decay_to_gs_zoom}
\end{figure}

In Fig. \ref{E1_ncap_decay_to_gs_zoom} we magnify the resonance structures
 in the range of $\epsilon_{\rm kin} = 5 - 7$ MeV.  We see clearly 
interferences between the resonances and the non-resonant capture. Note
that dominant partial wave contributing to the peak as well as the interference pattern are quite 
different for different resonances, reflecting the non-collective
nature of these resonance states.

\subsubsection{Transitions to the low-lying states $2^{+}_{1,2}$}\label{discuss_2plus}

The transitions decaying to the low-lying quadrupole states $2^{+}_{1,2}$ are most dominant
at low energy as seen in Fig. \ref{E1_ncap_total}.(red and blue curves).  Here we shall discuss the one for $2_1^{+}$
since the cross sections for $2_1^{+}$ and
$2_2^{+}$ behave similarly.

\begin{figure}
\centering
\begin{minipage}{\columnwidth}
\includegraphics[width=0.7\columnwidth]{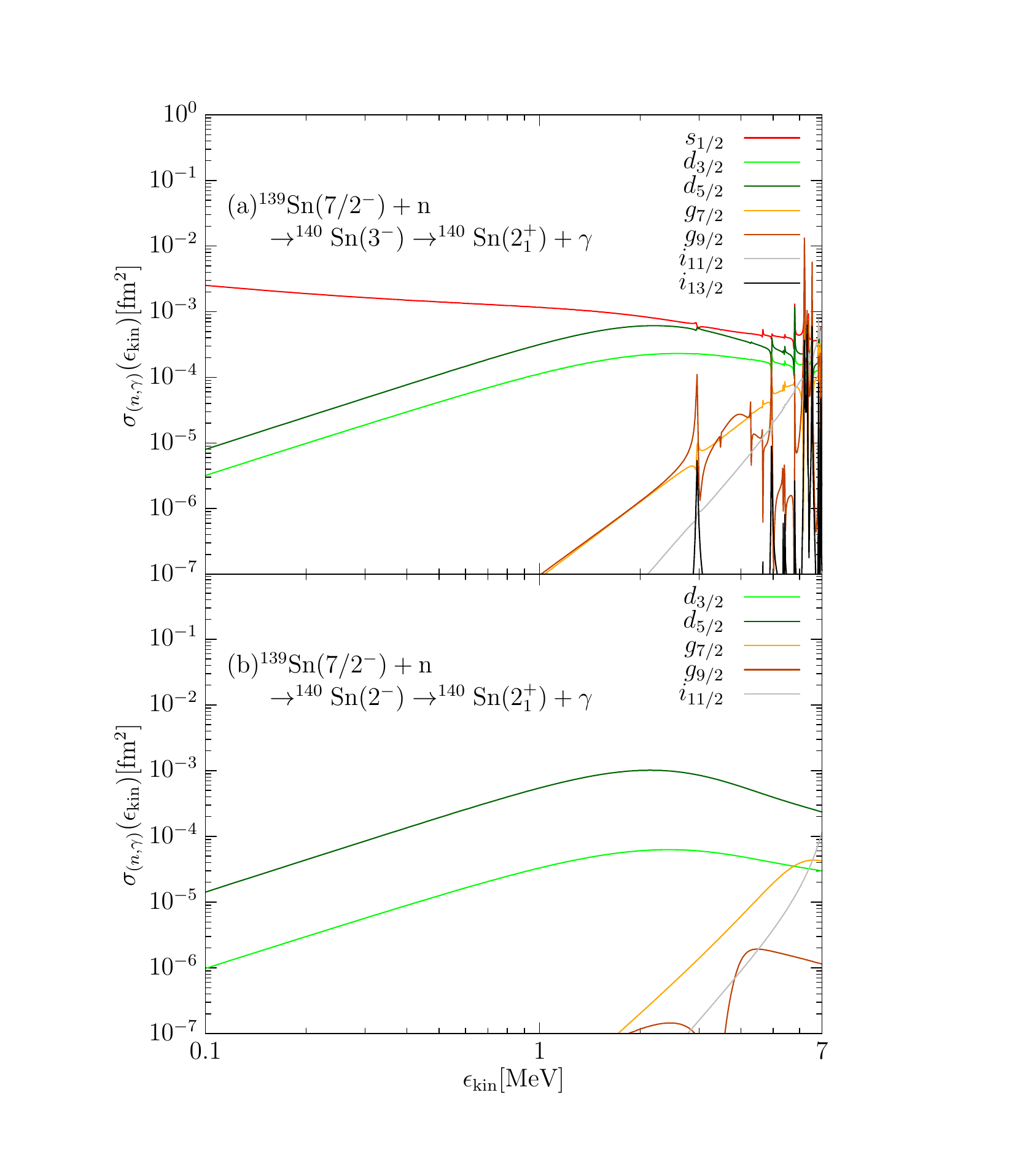}
\end{minipage} 
\caption{
The calculated partial $(n,\gamma)$ cross sections for
$^{139}{\rm Sn}({\small 7/2}^{-}) + n \to ^{140}{\rm Sn}(2^{-},3^{-}) \to ^{140}{\rm Sn}(2^{+}_{1}) + \gamma$ for E1 transitions,
plotted separately for different partial waves of the incident neutron; $s_{1/2}$ (red curve), $d_{3/2}$ (green),
and $d_{5/2}$ (dark green) etc..  The panel (a) is for the total spin $L^{\pi}=3^{-}$, and (b) 
for $L^{\pi}=2^{-}$.
The horizontal axis is the neutron kinetic energy $\epsilon_{\rm kin}$.
The smoothing parameter  is $\eta=10^{-5}$ MeV.
}
\label{E1_ncap_decay_to_1st2}
\end{figure}

The spin-parity of the total system  relevant to the present case is 
$L^{\pi}=1^{-}, 2^{-}$ and $3^{-}$.
Figure \ref{E1_ncap_decay_to_1st2} shows 
the partial cross section decomposed with respect to the partial waves 
$l_p j_p$ of the incident neutron, plotted separately 
for  $L^{\pi}=3^{-}$ and $2^{-}$ in panels (a) and (b), respectively.

It is seen from comparison of Fig. \ref{E1_ncap_total} and Fig. \ref{E1_ncap_decay_to_1st2}  that 
a large cross section at low energy is attributed to the $s$-wave capture.  
A noticeable feature is that 
 the $s$-wave capture is present only  for $L^{\pi}=3^{-}$ but not for $L^{\pi}=2^{-}$.
 Possible partial waves of incident neutron allowed for the total spin-parity $L^{\pi}=3^{-}$
 are $s,d,g$ and $i$  due to the  the angular momentum coupling to spin  $7/2^{-}$ of the target  
$^{139}{\rm Sn}$. However for $L^{\pi}=2^{-}$ allowed partial waves are $d,g$ and $i_{11/2}$ for $L^{\pi}=2^{-}$,
excluding the $s$-wave. This is the same reason why there is no $s$-wave capture in the
 transitions to the ground state.  Consequently the transitions to $2^{+}_{1,2}$  dominate  at low energy.
 
\begin{figure}
\centering
\begin{minipage}{\columnwidth}
\includegraphics[width=0.6\columnwidth]{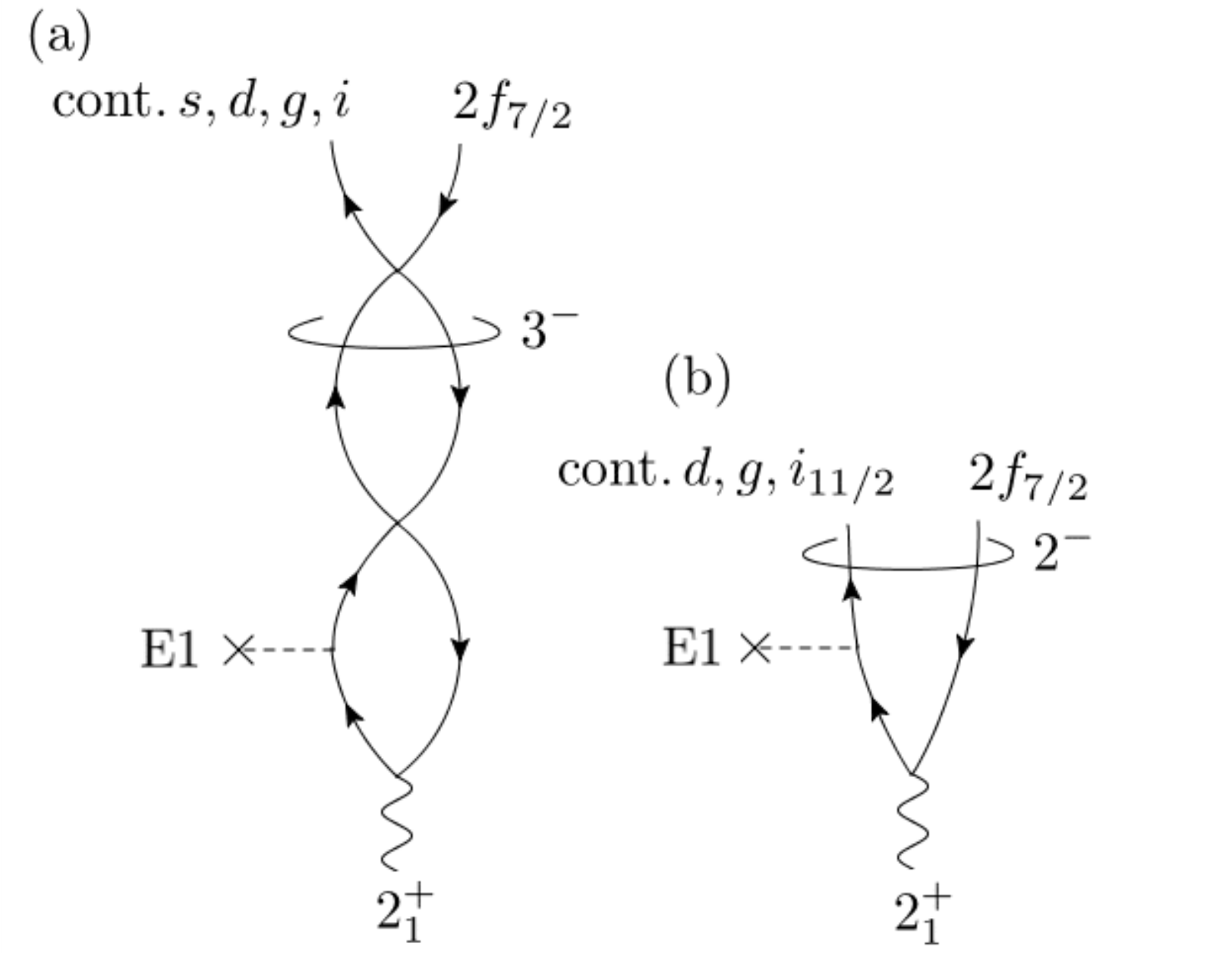}
\end{minipage} 
\caption{The diagrams representing $(n,\gamma)$ reaction of $^{139}{\rm Sn}({\small 7/2}^{-}) + n \to ^{140}{\rm Sn(2^{-},3^{-})} \to ^{140}{\rm Sn(2^{+}_{1}) + \gamma}$;   (a) for the total spin $L^{\pi}=3^{-}$ and (b)
for $L^{\pi}=2^{-}$.}
\label{diagram_1st2}
\end{figure}

Another significant feature is that many resonance-like structures with narrow width are seen for 
$L^{\pi}=3^{-}$, but not for $L^{\pi}=2^{-}$, though the impact on the absolute 
cross section is small.  Similarly to the transition to the ground state
 (cf the previous subsection), the narrow resonances
originate from the coupling between the non-resonant continuum states 
(  $s,d,g$ and $i$ wave coupled to $^{139}{\rm Sn}$ with $7/2^{-}$) and 
non-collective $1p1h$ configurations 
of both neutrons and protons, e.g. $\nu[(1h_{9/2})(3s_{1/2})^{-1}]$, $\nu[(2f_{5/2})(3s_{1/2})^{-1}]$, $\nu[(1h_{9/2})(2d_{3/2})^{-1}]$, $\nu[(3p_{3/2})(2d_{3/2})^{-1}]$, and $\nu[(2f_{5/2})(2d_{3/2})^{-1}]$ (cf.  Fig. \ref{diagram_1st2}(a)).
 However, the spin-independent 
residual interaction  Eq.(\ref{v_ph}) does not cause mixing among  $1p1h$ states with 
unnatural spin-parity $2^{-}$.  In this case only the
direct transition shown by the diagram Fig. \ref{diagram_1st2}(b) 
 is relevant and 
the non-collective $1p1h$ states does not
show up as resonances.

\subsubsection{Transitions to the octupole vibrational state $3^{-}_{1}$}\label{discuss_3minus}

\begin{figure}
\centering
\begin{minipage}{\columnwidth}
\includegraphics[width=0.7\columnwidth]{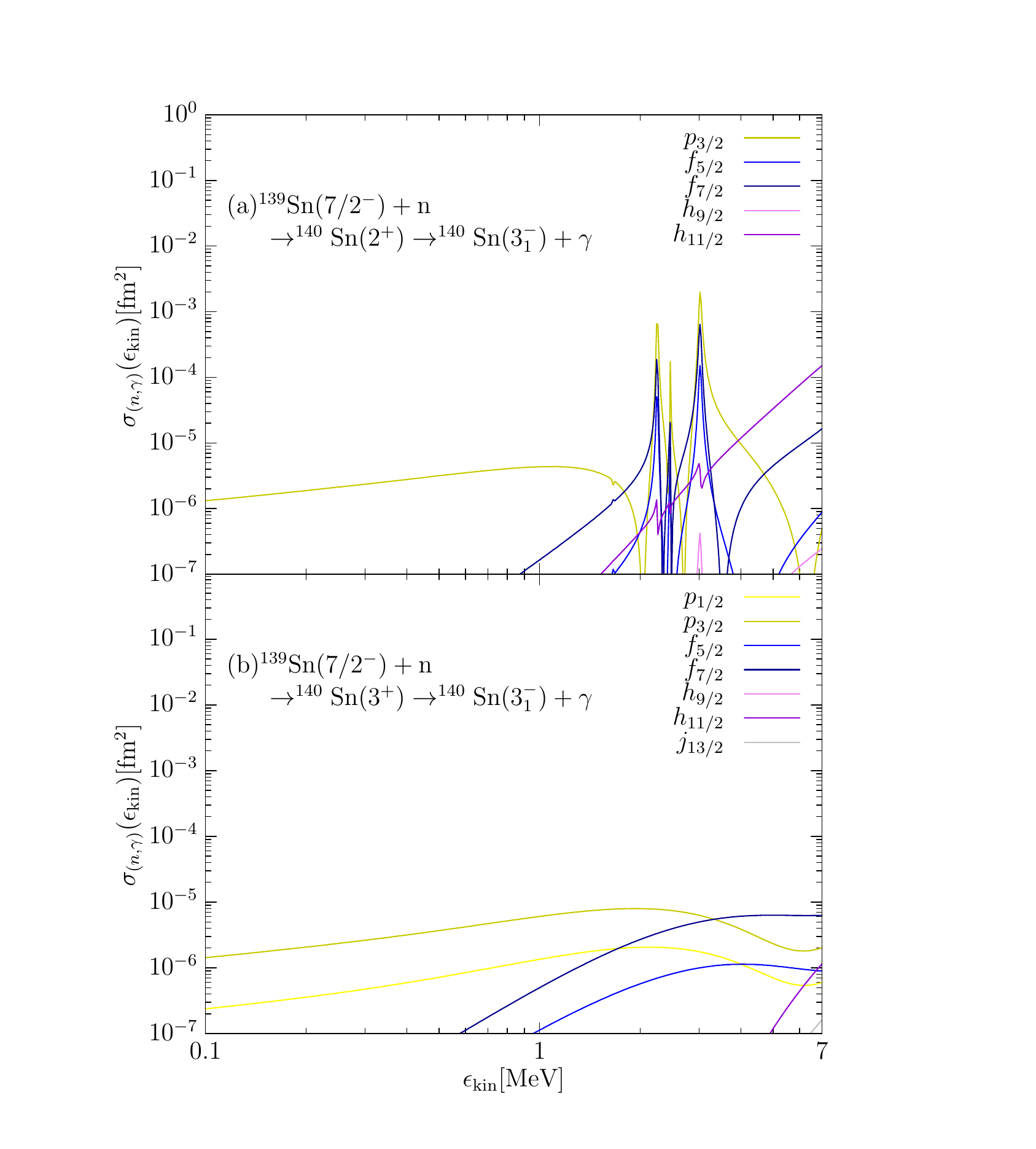}
\end{minipage} 
\caption{
The calculated partial $(n,\gamma)$ cross sections for
$^{139}{\rm Sn}({\small 7/2}^{-}) + n \to ^{140}{\rm Sn}(2^{+},3^{+}) \to ^{140}{\rm Sn}(3^{-}_{1}) + \gamma$ for E1 transitions,
plotted separately for different partial waves of the incident neutron; $p_{1/2}$ (yellow curve), $f_{5/2}$ (blue),
and $h_{9/2}$ (pink) etc..  The panel (a) is for the total quantum number $L^{\pi}=2^{+}$, and  (b)
for $L^{\pi}=3^{+}$.
The horizontal axis is the neutron kinetic energy $\epsilon_{\rm kin}$.
The smoothing parameter  is $\eta=10^{-5}$ MeV.
}
\label{E1_ncap_decay_to_1st3}
\end{figure}

Let us discuss the $(n,\gamma)$ reaction whose final state is the octupole vibrational state
$3^{-}_{1}$ at $E= 1.768$ MeV. The relevant spin-parity of the total system is $L^{\pi}=2^{+},3^{+}$ and $4^{+}$,
and we show in Fig. \ref{E1_ncap_decay_to_1st3} partial $(n,\gamma)$ cross sections for $L^{\pi}=2^{+}$ and $3^{+}$
separately. The low energy cross section $\epsilon_{\rm kin} \le 1$ MeV are dominated by the $p$-wave capture in both cases
since $l=1$ is the smallest orbital angular momentum of the neutron partial waves allowed both for  $L^{\pi}=2^{+}$ and $3^{+}$.
Note that the capture in the $p$-wave (and other negative-parity partial waves) would not be present 
if the collective octupole state did not exist  below the neutron separation energy. 

The cross section for $L^{\pi}=2^{+}$ (Fig. \ref{E1_ncap_decay_to_1st3}(a))
shows a significant
enhancement around $\epsilon_{\rm kin} \sim 2-3 $ MeV with resonant structures. These resonance behaviors originate from
excited $2^{+}$ states of $^{140}{\rm Sn}$ at $E=4.8$ and $5.6$ MeV, which appear
as relatively large peaks in the E2 and isoscalar quadrupole strength functions shown in Fig. 2(b) of Ref. \cite{Saito2021}.
These quadrupole states have some collectivity, i.e. consisting of
coherent neutron particle-hole admixture in addition to main proton particle-hole configurations
$\pi[(1g_{9/2})(1g_{9/2})^{-1}]$, and $\pi[(2d_{5/2})(1g_{9/2})^{-1}]$. We remark that the collectivity in the
quadrupole states is reflected in Fig. \ref{E1_ncap_decay_to_1st3}(a);
the neutron partial waves $p_{3/2}$ and $f_{7/2,5/2}$ contribute coherently to the resonance structures. Such a 
coherence effect is not seen in the narrow resonance structures shown in Figs. \ref{E1_ncap_decay_to_gs},
\ref{E1_ncap_decay_to_gs_zoom} and \ref{E1_ncap_decay_to_1st2} where the relevant resonant
states have a character of  non-collective $1p1h$ excitations.

\begin{figure}
\centering
\begin{minipage}{\columnwidth}
\includegraphics[width=0.6\columnwidth]{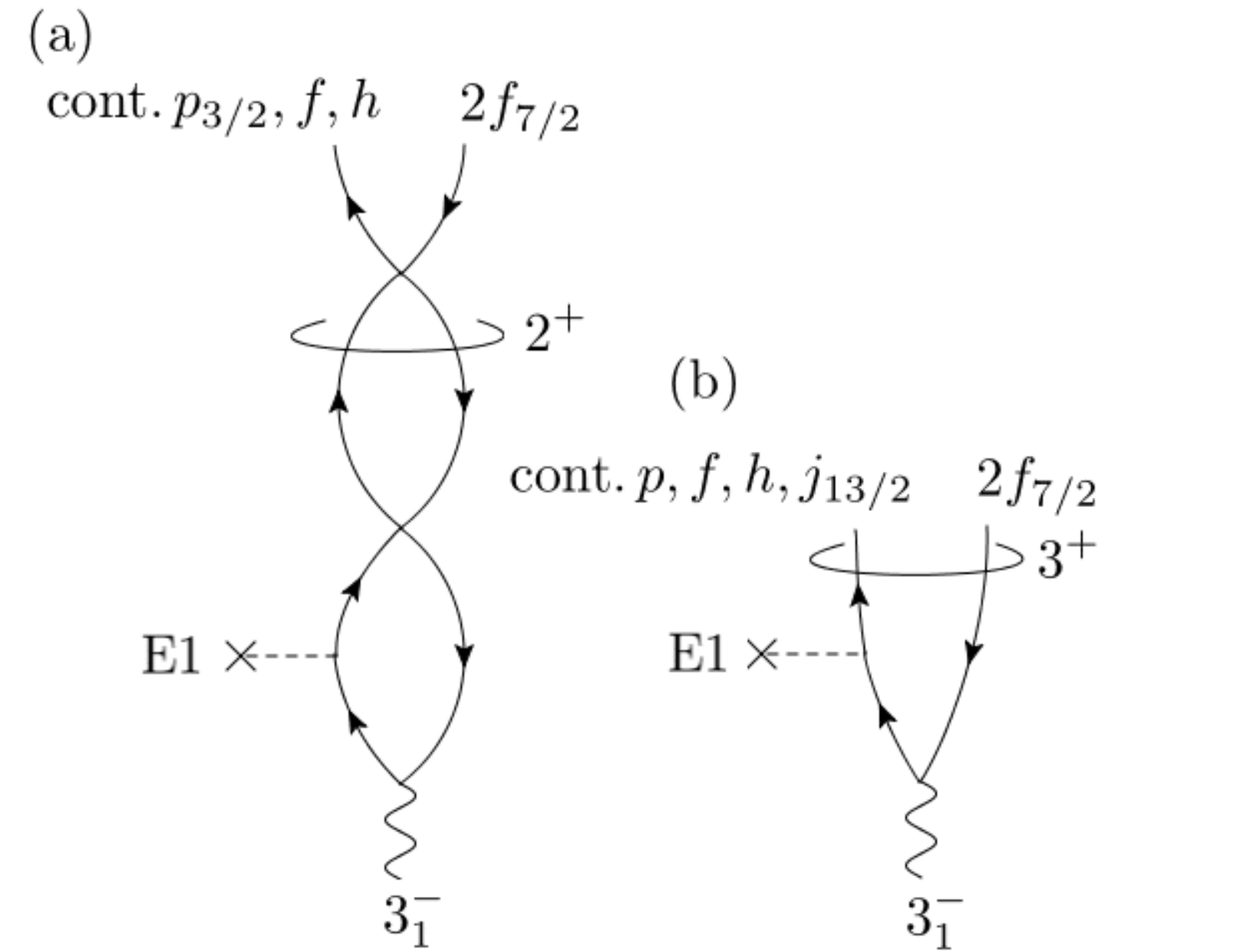}
\end{minipage} 
\caption{The diagrams representing $(n,\gamma)$ reaction of $^{139}{\rm Sn}({\small 7/2}^{-}) + n \to ^{140}{\rm Sn(2^{+},3^{+})} \to ^{140}{\rm Sn(3^{-}_{1}) + \gamma}$.}
\label{diagram_1st3}
\end{figure}

As for the case of $L^{\pi}=3^{+}$ (Fig. \ref{E1_ncap_decay_to_1st3}(b)), 
the energy dependence is smooth without narrow resonance structures 
 since there is no
effect of the correlation, similarly to Fig. \ref{E1_ncap_decay_to_1st2}(b). The process here is
the direct capture represented by the
diagram Fig. \ref{diagram_1st3}(b).

\subsection{Cross section at higher energy}\label{discuss_he}

Let us examine a global behavior of the calculated cross sections in a wide energy region covering up to $\epsilon_{\rm kin} \sim 20$ MeV 
although the result at high energies is not relevant to the r-process. We remark also 
that the model space of the present theory does not
include multi-particle-multi-hole configurations, needed to describe the high level density and the complex structure of compound states at 
high excitation energy.
In other words the present calculation lacks a part of the $(n,\gamma)$ processes
 that is usually modeled in terms of the
statistical treatment of the compound states.

\begin{figure}
\centering
\begin{minipage}{\columnwidth}
\includegraphics[width=0.52\columnwidth]{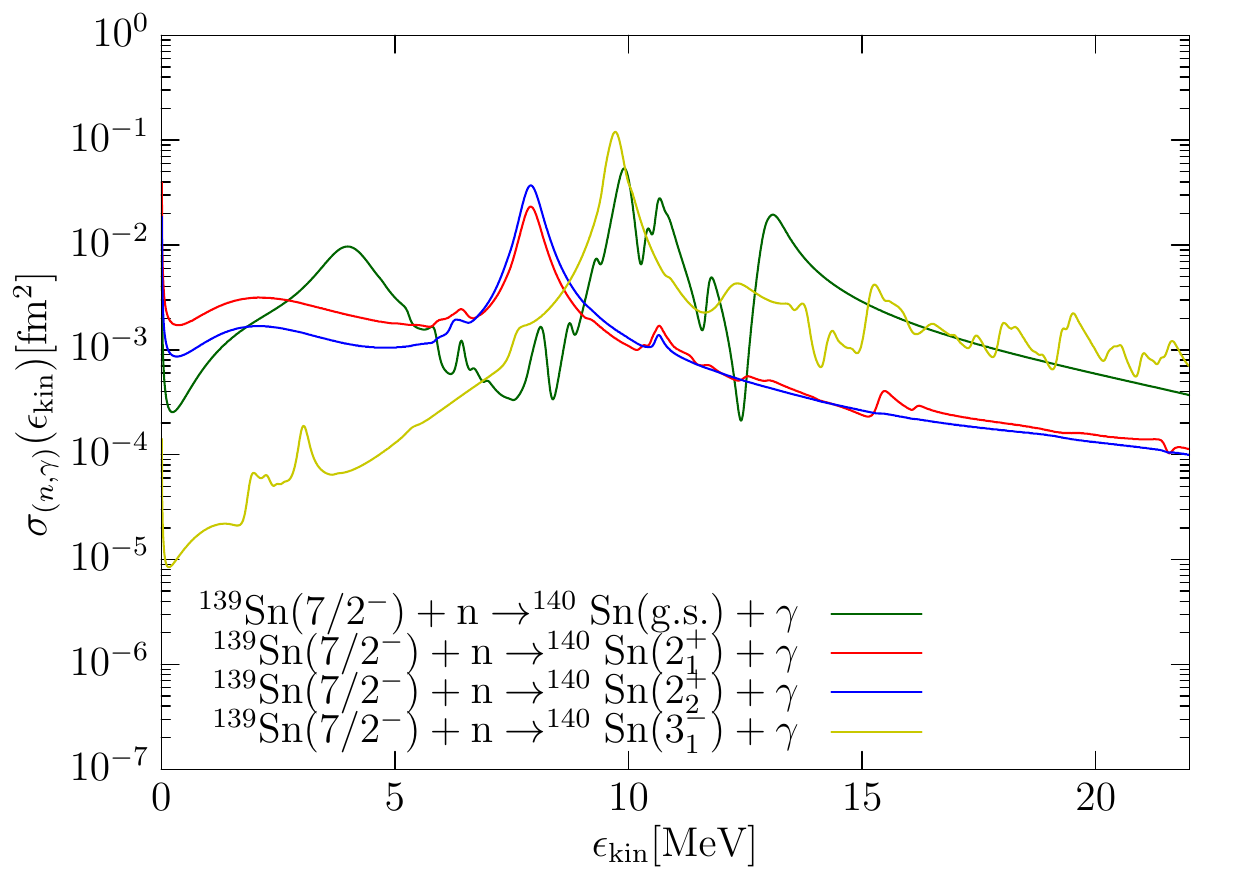}
\end{minipage} 
\caption{
The calculated $(n,\gamma)$ cross sections for $^{139}{\rm Sn}({\small 7/2}^{-})+ n \to ^{140}{\rm Sn}$
with  E1 transitions populating different low-lying states in $^{140}{\rm Sn}$; the ground state (plotted with a green curve),
the low-lying $2^{+}_{1}$ and $2^{+}_{2}$ states (red and blue curves respectively), and
the octupole vibrational state $3^{-}_{1}$ (yellow curve). 
The horizontal axis is the neutron kinetic energy $\epsilon_{\rm kin}$.  
The smoothing parameter  is $\eta=0.1$ MeV.}
\label{E1_ncap_total_HE}
\end{figure}

Figure \ref{E1_ncap_total_HE} shows the $(n,\gamma)$ cross section for transitions 
to the low-lying $2_{1,2}^{+}$ and $3_{1}^{-}$ states as well as the ground
state in $^{140}{\rm Sn}$.
It is the same as Figure \ref{E1_ncap_total} but for $\epsilon_{\rm kin} < 20$ MeV. The smoothing constant $\eta$ is set to $\eta=0.1$ MeV,
which washes out narrow resonances.

We see several resonance-like peaks  having sizable width at $\epsilon_{\rm kin} \approx 4$ MeV, 8 MeV, 10 MeV
and a bunch of peaks around $\epsilon_{\rm kin} \approx 10 -13$ MeV.
The large peak at $\epsilon_{\rm kin}  \approx 4$ MeV is the one which we already discussed in subsection \ref{discuss_gs}.
It is essentially a single-particle resonance in the partial wave $g_{9/2}$, from which a single-particle E1 transition to
the  $2f_{7/2}$ orbit occurs.

The peak at $\epsilon_{\rm kin}  \approx 8$ MeV seen in the transition to the $2^{+}_{1,2}$ states 
reflects the single-particle $i_{11/2}$ resonance. Note that this peak is seen only for the transtion
to the $2^{+}_{1,2}$ states, but not to the ground state. 
The single-particle E1 transition is possible from the resonant $i$ orbit  to the bound
$1h_{9/2}$ orbit (see Fig. \ref{diagram_1st2}(a)), which is occupied in the $2^{+}_{1,2}$ states 
as one of the main particle-hole configurations (see Table. \ref{tableXamp2}). 

The largest peak at $\epsilon_{\rm kin}  \approx 10$ MeV is present in the transition to the octupole state $3^{-}_{1}$.
It originates from a single-particle resonance in the partial wave  $j_{15/2}$.  
We remark that the collectivity of the low-lying octupole state $3^{-}_{1}$ plays an essential role. 
The single-particle $j_{15/2}$ resonance has sizable E1 matrix element for a transition to 
the neutron $1i_{13/2}$ orbit, 
which is, however, unbound (resonance)  in the present model. Therefore the capture is not possible 
via the single-particle process.   In contrast,
the   $3^{-}_{1}$ state with the collective character of
surface vibration can be a final state of the E1 transition from the 
neutron single-particle $j_{15/2}$ resonance 
since it contains configuration
$(1i_{13/2})(2f_{7/2})^{-1}$ as one of main $1p1h$ components (cf. Table \ref{tableXamp3}).
 
A distribution of several peaks around $\epsilon_{\rm kin} \approx 10 -13$ MeV seen in the transition to the ground state (green curve) is another feature
that  points to importance of the 
collectivity in the $(n,\gamma)$ reaction. These peaks are not related to the single-particle resonance, but 
they originate from the giant dipole resonance (GDR) in $^{140}{\rm Sn}$ 
with spin-parity $1^{-}$ distributed around excitation energy $E \approx 12-16 $ MeV. The capture proceeds through the
GDR, which enhances gamma decay to the ground state.

\subsection{$(n,\gamma)$ reaction with E2 transition}\label{discuss_E2}

\begin{figure}
\centering
\begin{minipage}{\columnwidth}
\includegraphics[width=0.52\columnwidth]{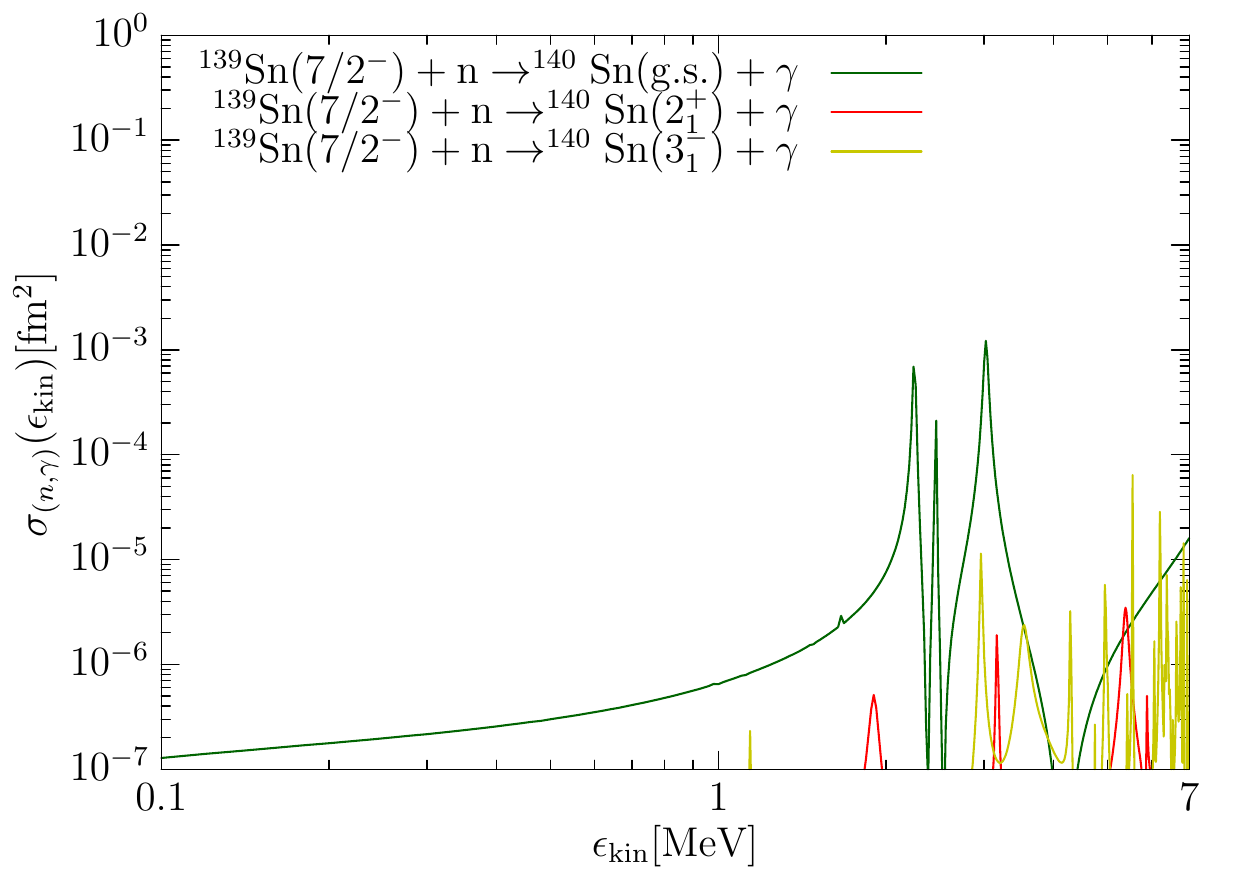}
\end{minipage} 
\caption{
The calculated $(n,\gamma)$ cross sections for $^{139}{\rm Sn}({\small 7/2}^{-})+ n \to ^{140}{\rm Sn}$
with  E2 transitions populating different low-lying states in $^{140}{\rm Sn}$; the ground state (plotted with a green curve),
the low-lying $2^{+}_{1}$  states (red curve), and
the octupole vibrational state $3^{-}_{1}$ (yellow curve). 
The horizontal axis is the neutron kinetic energy $\epsilon_{\rm kin}$.  
The cross section for $2^{+}_{2}$ is not seen as it is smaller than the plotted range.
The smoothing parameter  is $\eta=10^{-5}$ MeV.}
\label{E2_ncap_total}
\end{figure}

\begin{figure}
\centering
\begin{minipage}{\columnwidth}
\includegraphics[width=0.52\columnwidth]{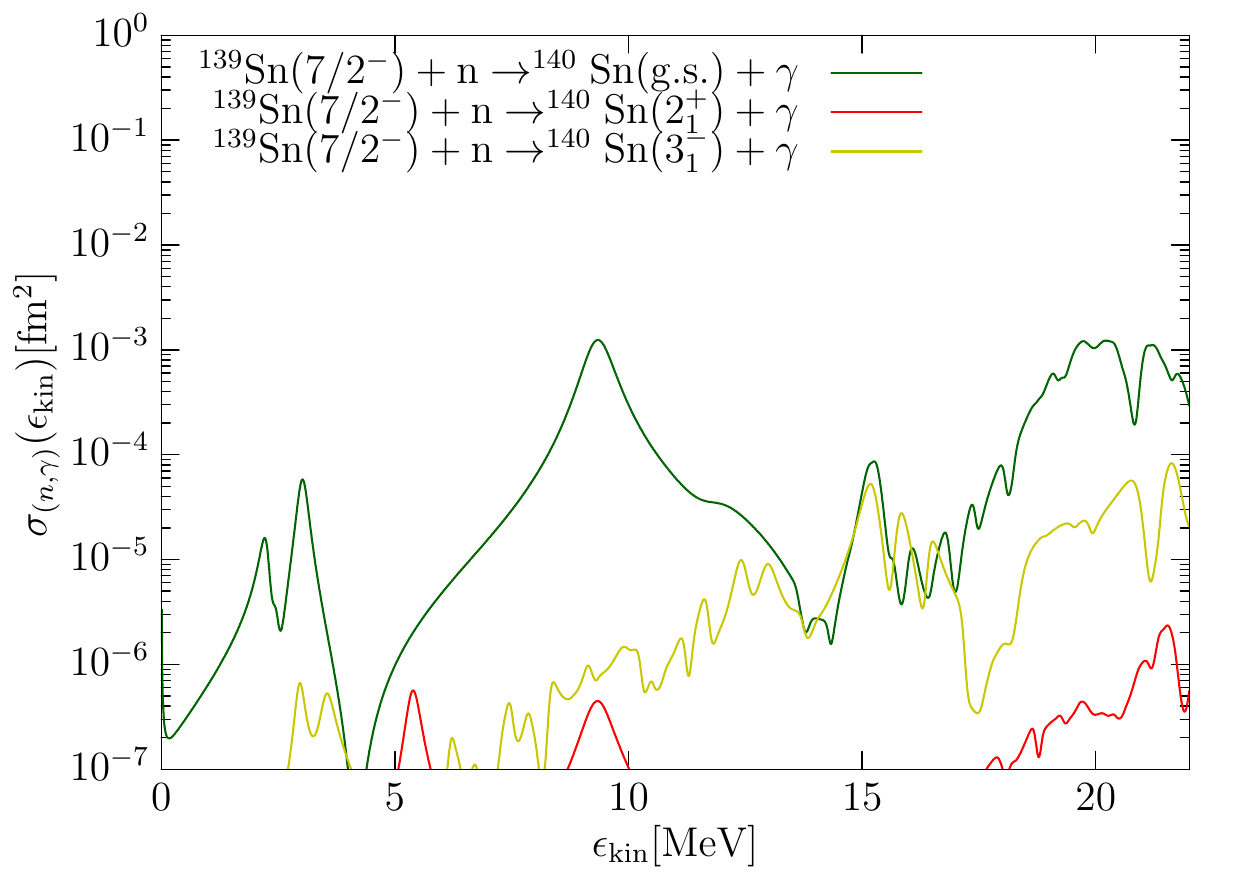}
\end{minipage} 
\caption{
The same as Fig. \ref{E2_ncap_total}, but for higher neutron kinetic energy
and larger smoothing parameter $\eta=0.1$ MeV.}
\label{E2_ncap_total_HE}
\end{figure}

We have calculated also the cross section for E2 transitions using the same formalism.
The obtained cross sections are shown in Fig. \ref{E2_ncap_total} and
\ref{E2_ncap_total_HE} for the low energy part and the high energy part, respectively.
The cross section for the E2 transition is smaller than those for E1 as is
seen by comparing with the results for the E1 case  (Figs. \ref{E1_ncap_total} and \ref{E1_ncap_total_HE}). 
An apparent reason is the kinematical factor which suppresses the
high multipole transitions.

A noticeable difference from the E1 case is that the transitions to the ground state is dominant 
over other transitions feeding the low-lying excited states $2_{1,2}^{+}$ and $3_{1}^{-}$.
Here we note that the adopted E2 operator acts only on protons, neglecting the recoil correction of
an order of $A^{-1}$.
For this reason both the continuum excited state and the low-lying excited states need to include proton  $1p1h$ configurations 
in order for the E2 transition to occur. Note that the calculated 
low-lying states in neutron-rich $^{140}{\rm Sn}$ has dominant neutron character
where the proton component is relatively small. This is especially the case for
 the $2^{+}$ states (cf. Table \ref{tableXamp2}). Consequently, the transition to the low-lying $2_{1,2}^{+}$ and $3_{1}^{-}$
 states are suppressed. For the transition to the ground state, on the contrary, 
 the E2 transition occurs 
 as far as the continuum excited states contains both proton $1p1h$ configurations and a scattering neutron orbit. 
 This is the case for the resonance peaks around $\epsilon_{\rm kin} = 2.2$ and $3.0$ MeV
 (corresponding to $2^{+}$ states  at excitation energy at $E=4.8$ and $5.6$ MeV), and 
 the broad peaks around $\epsilon_{\rm kin} \approx 10$ MeV and $\epsilon_{\rm kin} \approx 15-22$ MeV, which
 correspond to the ISGQR and IVGQR, respectively.
  
\section{Conclusion}
The continuum random-phase approximation (cRPA) with use of the Green's function in the coordinate-space representation enables us
to describe various types of correlated particle-hole excitations
and their coupling to  unbound single-particle orbits. Utilizing this feature of the cRPA, we have formulated a quantum many-body theory
to calculate the cross section of the
radiative  neutron capture reaction on neutron-rich nuclei with a scope of application to the r-process nucleosynthesis. 
We have given, in our previous publication \cite{Matsuo2015}, a formulation for the $(n,\gamma)$ reaction,
in which the final state of the gamma transition 
is restricted to the ground state of an even-even nucleus.
In the present study, we have introduced an extended formulation 
so that we can describe decay channels in which the gamma transition populates low-lying excited states.
With the cRPA approach it is possible to describe various excitation modes present in the scattering state
of $n + (A-1)$, including soft dipole excitation, the giant resonances as well as  non-collective excitations and
the single-particle resonances. Furthermore, we are able to takes into account  the correlation 
in the final states of the gamma transition, e.g. the low-lying quadrupole and
octupole vibrational states.   

In order to demonstrate the new features taken into account in the cRPA description of the $(n,\gamma)$ reaction, we have performed
numerical study of  the neutron capture  of neutron-rich nucleus $^{139}{\rm Sn}$ followed by E1 or E2 $\gamma$ transtions to the ground state or  low-lying excited states $2^{+}_{1},2^{+}_{2}$ and $3^{-}_{1}$ in $^{140}{\rm Sn}$. 
The RPA correlation  in the scattering states  produces 
resonance structures in the $(n,\gamma)$ cross section, which originates from particle-hole excitations with both collective
and non-collective characters. There appear resonances with rather wide
widths, arising from single-particle resonances and collective excitations such as the giant resonances. In addition
narrow resonances emerge from non-collective excitation modes, which are particle-hole excitations 
from bound to bound orbits of neutrons as well as those of protons.  
The correlation in the final states, i.e. the low-lying quadrupole and octupole states
$^{140}{\rm Sn}$, has also significant impact on the $(n,\gamma)$ cross section. 
A particular example is the transitions to the $3^{-}_{1}$ state, which is not present in
the single-particle model as this collective state appears below the neutron
separation energy only by including the RPA correlation. Furthermore additional neutron partial waves
contribute to the capture, compared with those in the single-particle model,
because of the strong configuration mixing of  many one-particle-one-hole states in the $3^{-}_{1}$ state.

Finally we  remark a scope for future developments of the present study. The pair correlation neglected in the present formulation
can be taken into account by a straightforward extension from the cRPA to the  continuum quasiparticle random-phase approximation (cQRPA). 
This will be done by combining the formulation in Ref. \cite{Matsuo2015} and that in the present study. This extension will makes it possible to describe
the $(n,\gamma)$ reaction systematically for open-shell spherical nuclei.  Extension of the model space to that beyond the RPA, including
many-particle many-hole states, is another direction of future development, which might be required for description of the neutron capture
in nuclei close to the stability line, e.g. in application to the s-process and capture reactions at high neutron energy.  
Description of the case where the final state nucleus  is an odd-$A$
or an  odd-odd nucleus also need to be studied in future.

\section{Acknowledgments}
The authors thank Kazuyuki Sekizawa and Kenichi Yoshida for valuable discussions. This work was
supported by the JSPS KAKENHI (Grant No. 20K03945).

\appendix
\section{$T$ matrix for $(\gamma,n)$ reaction}

We shall discuss a relation between the $T$ matrix $T^{\rm RPA}_{(\gamma, n)}=\bra{ph} \hat{V}_{\rm scf}(\hat{F}; \omega) \ket{0}$
given in the present cRPA theory  
and the $T$ matrix of the $(\gamma,n)$ reaction in the general theory of nuclear reaction.

Following the general reaction theory \cite{Taylor2006}, the $T$ matrix for the $(\gamma, n)$ reaction in the prior form is given by
\begin{align}
T_{(\gamma, n)} = \bra{\Psi^{(-)}_{n+(A-1)}(E)} \hat{M}\ket{\Psi_{i,A}},
\end{align}
\begin{align}
\ket{\Psi^{(-)}_{n+(A-1)}(E)} = \ket{\phi_{n} \Psi_{A-1}} + \hat{G}^{(-)}(E) \hat{V} \ket{\phi_{n} \Psi_{A-1}} ,
\end{align}
for the gamma-ray absorption caused by the multipole field $\hat{M}$ with a kinematical factor and the angular momentum omitted.
Here $\ket{\Psi_{i,A}}$ is the target nucleus $A$ at state $i$ and $\ket{\phi_{n} \Psi_{A-1}}$ is the final state in the $n+(A-1)$ channel consisting of an escaping neutron $\phi_{n}$ and the daughter nucleus $(A-1)$ (the state $\Psi_{A-1}$) . 
$\ket{\Psi^{(-)}_{n+(A-1)}(E)}$ is an energy eigenstate with an incoming boundary condition, which is
 given as a solution of the Lippman-Schwinger equation.
Here $\hat{V}$ is the interaction between the neutron and the 
subsystem $(A-1)$, and  
$\hat{G}^{(\pm)}(E)$ is a many-body Green's function
$
\hat{G}^{(\pm)}(E) = \frac{1}{E - \hat{H} \pm i\eta}
$
for the total Hamiltonian $\hat{H}=\hat{H}_0 + \hat{V}$ with $\hat{H}_0$ describing non-interacting  $n+(A-1)$ system. 
Note that the $T$ matrix can be written also as
\begin{align}
\label{t-n-gamma-post}
T_{(\gamma, n)} & = \bra{\phi_{n} \Psi_{A-1}} \hat{M}\ket{\Psi_{i,A}} 
 + \bra{\phi_{n} \Psi_{A-1}} \hat{V}\hat{G}^{(+)}(E)\hat{M}\ket{\Psi_{i,A}} 
 \end{align}
 which is the post form representation of the $T$ matrix. Besides the
 photo-nuclear interaction $\hat{M}$, the interaction $\hat{V}$ also contributes to the 
 $T$ matrix in the post form as represented by the second term in the r.h.s. of Eq.(\ref{t-n-gamma-post}).

The RPA $T$ matrix (c.f. Eqs.(\ref{def_Vscf}) and (\ref{matel_Vscf})) is written as
\begin{align}
\label{t-n-gamma-rpa}
T^{\rm RPA}_{(\gamma, n)}
  = \bra{ph} \hat{F} \ket{0} + \bra{ph} \hat{V}_{\rm ind}(\hat{F};\omega) \ket{0}
 \end{align}
in terms of the induced field 
 \begin{align}
 \hat{V}_{\rm ind}(\hat{F};\omega) = \int dx \frac{\delta \hat{U}(x)}{\delta \rho}\delta\rho(x,\omega), \ \ \ 
 \frac{\delta \hat{U}(x)}{\delta \rho}=\frac{\delta {U}(x)}{\delta \rho}\hat{\rho}(x).
 \end{align}
 
The first and second terms in Eq.(\ref{t-n-gamma-rpa}) correspond to the respective terms in Eq.(\ref{t-n-gamma-post}). To see this, 
we first note correspondence in the initial and final states:
\begin{align} 
\bra{ph} \leftrightarrow \bra{\phi_{n} \Psi_{A-1}}, \ \ \ 
\hat{O}_{i}^{\dagger} \ket{0} \leftrightarrow   \ket{\Psi_{i,A}}  
\end{align}
assuming the excited state $\ket{\Psi_{i,A}}$ of the nucleus $A$ is described by the RPA creation operator
$\hat{O}_{i}^{\dagger}$ acting on the ground state $\ket{0}$ in the static selfconsistent field.  
The particle-hole state $\ket{ph}=a_p^{\dagger}a_h\ket{0}$ corresponds to the scattering final state
$\ket{\phi_{n} \Psi_{A-1}} $ if we assign the daughter nucleus $\Psi_{A-1}$ as a one-hole
state $a_h \ket{0}$  whereas the particle $a_p^{\dagger}$ to the neutron scattering wave $\phi_{n}$.
We see then correspondence 
\begin{align}
\bra{ph}\hat{F}\ket{0}= \bra{ph} \hat{M} \ket{i}   \leftrightarrow 
\bra{\phi_{n} \Psi_{A-1}} \hat{M}\ket{\Psi_{i,A}} 
\end{align}
between the first terms in the r.h.s. of Eq.(\ref{t-n-gamma-rpa}) and (\ref{t-n-gamma-post}). See also Fig.\ref{Fdiagram}(a).

Concerning the second term of the RPA $T$ matrix, the induced field 
 \begin{align}
 \hat{V}_{\rm ind}(\omega) = \int dx \frac{\delta {U}(x)}{\delta \rho}\hat{\rho}(x)\delta\rho(x,\omega)
 \end{align}
is a mean-field part of the RPA residual interaction
$ \hat{V}_{\rm RPA} = \frac{1}{2} \int dx \frac{\delta {U}(x)}{\delta \rho}\hat{\rho}(x)\hat{\rho}(x)$. It is associated with 
the perturbation represented by the density response $\delta\rho(x,\omega)$, which is brought by the action 
$\hat{F}=[\hat{M},\hat{O}_{i}^\dagger]$ of the multipole field
$\hat{M}$ on the initial state $\hat{O}_i^\dagger\ket{0}$. We thus see a
correspondence
\begin{align}
\hat{V}_{\rm ind}(\hat{F}; \omega) \ket{0}
  \leftrightarrow 
\hat{V}\hat{G}^{(+)}(E)\hat{M} \ket{\Psi_{i,A}}
 \end{align}
 under an approximation that the interaction $\hat{V}$ is replaced by the RPA residual interaction
  $ \hat{V}_{\rm RPA}$ and the model space is limited to the one-particle-one-hole subspace. Thus we find that 
  the second terms in the r.h.s. of Eq.(\ref{t-n-gamma-rpa}) and (\ref{t-n-gamma-post})  corresponds to each other:
\begin{align}
 \bra{ph} \hat{V}_{\rm ind}(\hat{F}; \omega) \ket{0}
  \leftrightarrow 
 \bra{\phi_{n} \Psi_{A-1}} \hat{V}\hat{G}^{(+)}(E)\hat{M}\ket{\Psi_{i,A}}.
 \end{align}

It is noted also that the general form of the $T$ matrix and the RPA $T$ matrix have a similar
structure; the interaction $\hat{V}$ or $\hat{V}_{\rm RPA}$ are taken into account up to infinite order as
\begin{align}
 T_{(\gamma, n)} & = \bra{\phi_{n} \Psi_{A-1}} \left(\hat{M} + \hat{V}\hat{G}^{(+)}_0(E)\hat{M} 
 + \hat{V}\hat{G}^{(+)}_0(E)\hat{V}\hat{G}^{(+)}_0(E)\hat{M}+ \cdots\right) \ket{\Psi_{i,A}} ,
 \end{align}
 \begin{align}
   T_{(\gamma, n)}^{\rm RPA} & = \bra{ph} \left( \hat{F} +
   \frac{\delta \hat{U}}{\delta \rho} {R}_0(\omega) {F} +   \frac{\delta \hat{U}}{\delta \rho} {R}_0(\omega)  \frac{\delta {U}}{\delta \rho} {R}_0(\omega) {F} + \cdots
  \right) \ket{0}
  \end{align}
  where the density response $\delta\rho(x,\omega)$ obeying the linear response equation (\ref{linear_response_equation_delta_rho}) is symbolically represented represented as Eq.(\ref{infinite_series_delta_rho}).
  Here  $\hat{G}^{(+)}_0(E)=\frac{1}{E-\hat{H}_0+i\eta}$ is the unperturbed Green's function whereas ${R}_0(\omega)$ is
  the unperturbed response function Eq.(\ref{R_0_spectrum}).
   A diagramatic representation is shown in Fig.\ref{Fdiagram}.

\section{The $(\gamma,n)$ cross section under spherical symmetry}

The self-consistent field $\hat{V}^{\rm scf}_{LM}(\omega) = \hat{F}_{LM} + \hat{V}^{\rm ind}_{LM}(\omega)$ is a rank $L$ non-local one-body operator.
The radial coordinate representation of this matrix element is expressed as
\begin{align}
\label{radial_v_scf}
v^{{\rm scf}}_{L, l^{'}j^{'},lj}(r_{x}, r_{y}, \omega) = F_{L, l^{'}j^{'},lj}(r_{x}, r_{y}) + \bra{l^{'}j^{'}} |Y_{L}| \ket{lj} \frac{\delta U}{\delta \rho}(r_{x}) \frac{1}{r^{2}_{x}} \delta \rho_{L}(r_{x}, \omega) \delta(r_{x} - r_{y}).
\end{align}
The radial matrix element $F_{L, l^{'}j^{'},lj}(r_{x}, r_{y}) $ of the operator $\hat{F}_{LM}$ is given in Eq.(34) of Ref. \cite{Saito2021}. 

Inserting Eq.(\ref{radial_v_scf}) into Eq.(\ref{neutron_emi_cross_section_radial}), we obtain
\begin{align}
&\sigma^{\lambda}_{iL_{i} + \gamma\to [\epsilon l_{p}j_{p}h]_{L}}(E_{\gamma}) = - \frac{f(E_{\gamma})}{\pi(2L_{i} + 1)} \notag \\
&\times {\rm Im} \Big\{ |\bra{l_{p}j_{p}} |Y_{L}| \ket{l_{h}j_{h}}|^{2} \iint dr_{x} dr_{x^{'}} \phi^{*}_{n_{h}l_{h}j_{h}}(r_{x}) \frac{\delta U}{\delta \rho}(r_{x}) \frac{1}{r^{2}_{x}} \delta \rho^{*}_{L}(r_{x}, \omega_{\gamma} + \omega_{i}) \notag \\
&\quad \times G_{0c, l_{p}j_{p}}(r_{x}, r_{x^{'}}, \epsilon_{h} + \hbar \omega_{\gamma} + \hbar \omega_{i} + i \eta) \frac{\delta U}{\delta \rho}(r_{x^{'}}) \frac{1}{r^{2}_{x^{'}}} \delta \rho_{L}(r_{x^{'}}, \omega_{\gamma} + \omega_{i})  \phi_{n_{h}l_{h}j_{h}}(r_{x^{'}}) \notag \\
&+ \bra{l_{p}j_{p}} |Y_{L}| \ket{l_{h}j_{h}}^{*} \iiint dr_{x} dr_{x^{'}} dr_{y^{'}} \phi^{*}_{n_{h}l_{h}j_{h}}(r_{x}) \frac{\delta U}{\delta \rho}(r_{x}) \frac{1}{r^{2}_{x}} \delta \rho^{*}_{L}(r_{x}, \omega_{\gamma} + \omega_{i}) \notag \\
&\quad \times G_{0c, l_{p}j_{p}}(r_{x}, r_{x^{'}}, \epsilon_{h} + \hbar \omega_{\gamma} + \hbar \omega_{i} + i \eta) F_{L, l_{p}j_{p}, l_{h}j_{h}}(r_{x^{'}}, r_{y^{'}}) \phi_{n_{h}l_{h}j_{h}}(r_{y^{'}}) \notag \\
&+ \bra{l_{p}j_{p}} |Y_{L}| \ket{l_{h}j_{h}} \iiint dr_{x} dr_{y} dr_{x^{'}} \phi^{*}_{n_{h}l_{h}j_{h}}(r_{y}) F^{*}_{L, l_{p}j_{p}, l_{h}j_{h}}(r_{x}, r_{y}) \notag \\
&\quad \times G_{0c, l_{p}j_{p}}(r_{x}, r_{x^{'}}, \epsilon_{h} + \hbar \omega_{\gamma} + \hbar \omega_{i} + i \eta) \frac{\delta U}{\delta \rho}(r_{x^{'}}) \frac{1}{r^{2}_{x^{'}}} \delta \rho_{L}(r_{x^{'}}, \omega_{\gamma} + \omega_{i}) \phi_{n_{h}l_{h}j_{h}}(r_{x^{'}}) \notag \\
&+ \iiiint dr_{x} dr_{y} dr_{x^{'}} dr_{y^{'}} \phi^{*}_{n_{h}l_{h}j_{h}}(r_{y}) F^{*}_{L, l_{p}j_{p}, l_{h}j_{h}}(r_{x}, r_{y}) \notag \\
&\quad \times G_{0c, l_{p}j_{p}}(r_{x}, r_{x^{'}}, \epsilon_{h} + \hbar \omega_{\gamma} + \hbar \omega_{i} + i \eta) F_{L, l_{p}j_{p}, l_{h}j_{h}}(r_{x^{'}}, r_{y^{'}}) \phi_{n_{h}l_{h}j_{h}}(r_{y^{'}}) \Big\} .
\end{align}
The function $G_{0c,l_{p}j_{p}}(r_{x}, r_{x^{'}}, \epsilon)$ is the single-particle Green's function $G_{0c}$ in the
partial wave $l_p j_p$. It is calculated exactly using Eq.(B2) of Ref. \cite{Saito2021}, together with the subtraction of the bound orbits, see Eq.(\ref{G_0c}).
The correlation in the low-lying state $\ket{iL_iM_i}$ is reflected in $ F_{L, l^{'}j^{'},lj}$ while the correlation
in the continuum RPA states $\ket{kLM(E)}$ is in the density flucutuation $\delta\rho_L$.

\bibliography{extended_DC_refs}

\end{document}